\documentclass[11pt]{article}

\usepackage{mathrsfs}
\usepackage{verbatim}
\usepackage{amsmath}
\usepackage{amsfonts}
\usepackage{amssymb}
\usepackage{latexsym}
\usepackage{graphicx}
\usepackage[nocompress]{cite}
\usepackage{hyperref}
\usepackage{xcolor}
\usepackage{comment}
\newtheorem{theorem}{\bf Theorem}[section]
\newtheorem{lemma}[theorem]{\bf Lemma}
\newtheorem{proposition}[theorem]{\bf Proposition}
\newtheorem{corollary}[theorem]{\bf Corollary}

\newenvironment{proof}{\noindent{\em Proof:}}{\quad \hfill$\Box$\vspace{2ex}}

\newenvironment{remark}[1][]{%
  \par\noindent\textbf{Remark%
    \if\relax\detokenize{#1}\relax.\else\ (#1).\fi}\ }%
  {\par\vspace{2ex}}

\numberwithin{equation}{section}
\numberwithin{figure}{section}
\numberwithin{table}{section}

\def \bR {{\mathbb R}}
\def \cA {{\mathcal A}}
\def \cB {{\mathcal B}}
\def \cE {{\mathcal E}}

\DeclareMathOperator{\sinc}{sinc}

\def \qquad {\quad\quad}

\makeatletter

\newcommand{\Rmnum}[1]{\expandafter\@slowromancap\romannumeral #1@}
\makeatother

\begin{document}
	
	\title{Sharp Rates for Optimal Recovery of Bandlimited Functions from Localized Oversampling}
	\author{Wentao Huang\thanks{Corresponding author. Department of Mathematics, The University of Hong Kong, Hong Kong, P.R. China. E-mail address: {\textit{huangwt@hku.hk}}.},
\qquad
    Haizhang Zhang\thanks{School of Mathematics, Sun Yat-sen University, Zhuhai, P.R. China. E-mail address: {\textit{zhhaizh2@sysu.edu.cn}}. Supported in part by the National Natural Science Foundation of China under grants  12671642 and 12371103.}}
	\date{}
	\maketitle
    \begin{abstract}
		It is well-known that one can recover a function $f$ bandlimited to $[-\delta,\delta]$ ( $0<\delta<\pi$) from its finite oversampling data $f(j),-n+1\le j\le n$, with an approximation error exponentially decaying to zero as $n$ tends to infinity. Various regularized Shannon sampling formulas have been proposed to achieve such exponential approximation. A recent sinh regularized formula achieved so far the best rate
        \[
        O\biggl(\frac1{\sqrt{n}}\exp(-(\pi-\delta)n)\biggr).
        \]
        We establish that the optimal recovery algorithm in the worst-case scenario achieves an approximation error $\rho_{n,\delta}$ satisfying
        \[
		 \frac{c_*}{\sqrt{n}}(\tan\frac\delta 4)^{2n+1/2}\le \rho_{n,\delta}\le  \frac{C_*}{\sqrt{n}}(\tan\frac\delta 4)^{2n+1/2},
		\quad 0<\delta\leq\pi,\quad n\geq1,
        \]
        where $c_*,C_*$ are explicit positive constants independent of $n$ and $\delta$. This result identifies the best possible exponential rate to reconstruct bandlimited functions from two-sided finite oversampling data. In other words,
        \[
        \rho_{n,\delta} \asymp \frac{\sqrt{\tan\frac\delta 4}}{\sqrt{n}}\exp\bigl(-c_\delta(\pi-\delta)n\bigr),
        \]
        where $c_\delta$ is strictly decreasing on $(0,\pi]$ and
        \[
        \lim_{\delta\to\pi^-}c_\delta=1,\ \lim_{\delta\to 0^+}c_\delta=+\infty.
        \]
       For one-sided sample data $f(j)$, $-n+1\le j\le 0$, we also obtain the sharp approximation order of optimally recovering function values on $[0,1]$:
        \[
        \rho_{n,\delta}^-\asymp \sqrt{\delta/\pi}\sin^n(\delta/2),\ 0<\delta<\pi,\ n\ge1,
        \]
        which improves the celebrated result of H. J. Landau on extrapolation of bandlimited functions.
	\end{abstract}
	
	

	\section{Introduction}
	The Whittaker--Kotel'nikov--Shannon sampling theorem \cite{Shannon-1949} states that a
	bandlimited function can be completely recovered from its infinite values sampled at the Nyquist rate.
	In applications, only finitely many samples near the point of interest
	are available. The resulting question is one of optimal recovery
	\cite{Micchelli-Rivlin-survey}: how accurately can a bandlimited function be
	recovered from these prescribed finite data in the worst-case scenario?
    Its answer characterizes the best accuracy attainable from these localized samples and provides a  benchmark for sampling algorithms.

	The relation between finite observation and frequency limitation is
	central to classical time--frequency concentration theory. Slepian and
	Pollak \cite{Slepian-Pollak-1961} developed its description through
	prolate spheroidal wave functions, whose asymptotic behavior was studied
	by Slepian \cite{Slepian-1965}. His subsequent discussions of bandwidth,
	uncertainty, and modeling \cite{Slepian-1976,Slepian-1983} place these
	questions in a broader framework. Here we study the information
	provided by finitely many point samples, with the optimal recovery
	error as the quantity to be estimated.

	More precisely, we use the unitary Fourier transform and the
	Paley--Wiener space
	\[
	\widehat f(\omega)=\frac{1}{\sqrt{2\pi}}
	\int_{\mathbb R}f(t)e^{-it\omega}\,dt,
	\qquad
	\mathcal B_\delta
	:=\left\{f\in  L^2(\mathbb R)\cap C(\bR):
	\operatorname{supp}\widehat f\subseteq[-\delta,\delta]\right\},
	\]
	where \(0\leq\delta\leq\pi\). For an integer \(n\geq1\), set
	\(J_n:=\{-n+1,\ldots,n\}\). We desire to recover \(f\) on \([0,1]\)
	from \(I_n(f):=((j,f(j)):j\in J_n)\), with \(f\) ranging over the unit ball of \(\mathcal B_\delta\) defined by
    \[
    U(\mathcal B_\delta)=\{f\in \mathcal B_\delta:\|f\|_{L^2(\mathbb R)}\le 1\}.
    \]
    For a reconstruction method $\cA$ mapping $I_n(f)$ to a bounded function $\cA(I_n(f))$ on $[0,1]$, we measure its worst-case uniform error by
    \[
    \cE(\cA)=\sup_{f\in U(\cB_\delta)}\sup_{x\in[0,1]}\Bigl|f(x)-\cA\bigl(I_n(f)\bigr)(x)\Bigr|.
    \]
    A regularized Shannon sampling formula has the following form 
    \[
    \cA\bigl(I_n(f)\bigr)(x)=\sum_{j\in J_n}f(j)\sinc(x-j)\phi_{n,\delta}(x-j),
    \]
    where $\sinc(x)=\frac{\sin\pi x}{\pi x}$ for $x\ne0$, with $\sinc(0)=1$, and $\phi_{n,\delta}$ is a regularization function whose parameters may depend on $n$ and $\delta$.

     Various  reconstruction methods by regularized Shannon sampling formula with exponentially decaying approximation errors have been proposed in the literature. In the following, fix $0<\delta<\pi$ and note that the constants in the estimates may depend on $\delta$ but not on the function $f$ or number of sampling points $n$. 
    \begin{itemize}
        \item $\sinc$ regularization \cite[1966]{Jagerman}: with an integer $m$ depending on $n$ and $\delta$,  
			\[
			\cA_1(I_n
			f)(x):=\sum_{j\in J_n}f(j)\sinc(x-j)\sinc^{m}\biggl(\frac{\pi-\delta}{m}(x-j)\biggr),
			\]
            satisfies 
			\begin{equation}\label{sincbound}
			    \cE(\cA_1)\le
                \frac{C}{n(\pi-\delta)}\exp(-\frac{\pi-\delta}{e}n).
			\end{equation} 
            \item Gaussian regularization \cite[2003]{Qian1}, \cite[2006]{Qian2}, \cite[2009]{Micchelli-Xu-Zhang}, \cite[2019]{Chen-Zhang-Gaussian}, \cite[2025]{Yang-Zhang-nonuniform}: with a parameter $r>0$ depending on $n$ and $\delta$, 
            \[
			\cA_2(I_n
			f)(x):=\sum_{j\in J_n}f(j)\sinc(x-j)\exp\biggl(-\frac{(x-j)^2}{2r^2}\biggr),
			\]
		satisfies
			\begin{equation}\label{gaussianbound}
            \cE(\cA_2)\le
			\frac C{\sqrt{n}}\exp(-\frac{\pi-\delta}{2}n).
			\end{equation}
            \item $\sinh$ regularization \cite[2022]{Kircheis-Potts-Tasche}, \cite[2025]{Kircheis2025}, \cite[2026]{Jiang2026}, \cite[2026]{Chen2026}: with a parameter $\beta$ depending on $n$ and $\delta$, 
            \[
			\cA_3(I_n
			f)(x):=\sum_{j\in J_n}f(j)\sinc(x-j)\frac{1}{\sinh(\beta)}\sinh\biggl(\beta\sqrt{1-\frac{(x-j)^2}{n^2}}\biggr),
			\]
		satisfies
			\begin{equation}\label{sinhbound} 
            \cE(\cA_3)\le
			\frac C{\sqrt{n}}\exp(-(\pi-\delta)n).
			\end{equation}
    \end{itemize}

    These results improve the exponential decay of prescribed reconstruction formulas. The optimal reconstruction algorithm $\cA_{\mathrm{opt}}$, which minimizes the error among all recovery methods, follows from classical optimal recovery theory \cite{Micchelli-Rivlin-survey}. In fact, with
	$
    K_\delta(x,y)=\frac{\sin \delta(x-y)}{\pi(x-y)},\quad x\ne y,
    $
    and $K_\delta(x,x)=\delta/\pi$,
    	\[
		\cA_{\mathrm{opt}}(I_n(f))=\sum_{j\in J_n}c_j K_\delta(j,\cdot),
		\]
		where the coefficients $c_j$ are determined by
		\[
		\sum_{k\in J_n} c_kK_\delta(j,k)=f(j), \ \ j\in J_n.
		\]
    Note that as the matrix $[K_\delta(j,k):j,k\in J_n]$ is highly ill-conditioned, this optimal algorithm is not computationally tractable. The error $\rho_{n,\delta}:=\cE(\cA_{\mathrm{opt}})$ provides a lower benchmark for all regularized Shannon sampling formulas and all tractable reconstruction methods. By \cite{Micchelli-Xu-Zhang},
    \[
    \rho_{n,\delta}=\sup_{x\in[0,1]}\rho_{n,\delta}(x),
    \]
    where, with \(E_n:=\operatorname{span}\{e^{-ij\omega}:j\in J_n\}\), the pointwise error $\rho_{n,\delta}(x)$ is given by
    \begin{equation}
        \rho_{n,\delta}(x)
        =\frac{1}{\sqrt{2\pi}}
        \operatorname{dist}_{L^2([-\delta,\delta])}
        \bigl(e^{-ix\omega},E_n\bigr)
        =\frac{1}{\sqrt{2\pi}}
        \inf_{p\in E_n}
        \bigl\|e^{-ix\omega}-p\bigr\|_{L^2([-\delta,\delta])}.
        \label{eq:optimal learning rate}
    \end{equation}

    Understanding the sharp rate of the optimal error $\rho_{n,\delta}$ is a 
    fundamental question in sampling theory \cite{Micchelli-Xu-Zhang}. The 
    lower bound in \cite[Theorem~4.12]{Micchelli-Xu-Zhang}, together with 
    the sinh regularization introduced above, indicates that for some positive constants $C_1,C_2$,
	\[
	C_1\frac{\delta}{2n+1}
	\left(\frac{\delta}{4}\right)^{2n}
	\leq\rho_{n,\delta}\leq
	\frac{C_2}{\sqrt{n}}
	\exp(-(\pi-\delta)n),\ \ 0<\delta\le \pi.
	\]
    The main objective of this paper is to resolve this open question by showing in Theorem~\ref{theorem:sharp rate} of Section 2 that 
    \[
     \rho_{n,\delta}\asymp \frac{1}{\sqrt{n}}\tan^{2n+1/2}\!\bigl(\frac{\delta}{4}\bigr)
    \]
    in the sense that there are positive constants $c_*\leq C_*$, independent of $n$ and $\delta$, for which 
	\[
		 \frac{c_*}{\sqrt{n}}\tan^{2n+1/2}\frac{\delta}{4}\le \rho_{n,\delta}\le \frac{ C_*}{\sqrt{n}}\tan^{2n+1/2}\frac{\delta}{4},
		\quad 0<\delta\leq\pi,\quad n\geq1.
    \]
    These bounds retain the exact geometric factor
    \(\tan^{2n}(\delta/4)\), the algebraic prefactor \(n^{-1/2}\),
    and the small-bandwidth factor \(\sqrt{\tan(\delta/4)}\), with
    the same comparison constants for every \(n\geq1\) and
    \(0<\delta\leq\pi\), including the critical bandwidth $\delta=\pi$.
    Write
    \[
    \tan^{2n}\frac{\delta}{4}=\exp\!\Bigl(2n\log\bigl(\tan(\delta/4)\bigr)\Bigr)
    \]
    and compute that $\frac{\log\bigl(\tan(\delta/4)\bigr)}{\pi-\delta}$ is strictly increasing on $(0,\pi)$ with
    \[
    \lim_{\delta\to\pi^{-}}\frac{\log\bigl(\tan(\delta/4)\bigr)}{\pi-\delta}=-\frac12,\qquad \lim_{\delta\to 0^+}\frac{\log\bigl(\tan(\delta/4)\bigr)}{\pi-\delta}=-\infty.
    \]
    Define
    \[
    c_\delta:=-\frac{2\log(\tan(\delta/4))}{\pi-\delta},
    \qquad 0<\delta<\pi,
    \qquad c_\pi:=1.
    \]
    Then
     \[
        \rho_{n,\delta} \asymp \frac{\sqrt{\tan(\delta/4)}}{\sqrt{n}}\exp\bigl(-c_\delta(\pi-\delta)n\bigr),
        \]
        where $c_\delta$ is strictly decreasing on $(0,\pi]$ and
        \[
        \lim_{\delta\to\pi^-}c_\delta=1,\qquad \lim_{\delta\to 0^+}c_\delta=+\infty.
        \]
    This means that, for every fixed \(0<\delta<\pi\), the error bounds
    \eqref{sincbound}--\eqref{sinhbound} for regularized Shannon
    sampling formulas decay at strictly slower exponential rates
    than the optimal recovery error. In particular, for small bandwidths,
    these bounds are far from attaining the optimal exponential rate.

    The second objective of the paper concerns recovering the values of a 
    function $f\in \cB_\delta$ ($0<\delta<\pi$) on $[0,1]$ from one-sided 
    sampling data $I_n^{-}(f)=((j,f(j)):-n+1\le j\le0)$. The completeness of
    $\operatorname{span}\{e^{ikt}:k\ge0\}$ in $L^2([-\delta,\delta])$
    for $\delta<\pi$ \cite[p.~116, Corollary]{Young} shows that infinite
    one-sided data determine $f$. The extrapolation problem of recovery from past samples has a substantial history.
    Slepian \cite[Section~3.2]{Slepian-discrete-1978} studied the optimal
    one-step prediction error and its fixed-bandwidth geometric rate.
    Mugler and Splettst{\"o}sser
    \cite[Theorem~2]{Mugler-Splettstosser-1986} developed prediction formulas
    based on modified differences for arbitrary fixed real target positions,
    under the stronger oversampling condition \(0<\delta<\pi/2\).
    Landau \cite[Theorem~1 and Corollary]{Landau-extrapolation-1986}
    established fixed-bandwidth exponential recovery on extrapolation
    intervals whose lengths grow linearly with the number of samples.
    Vaidyanathan \cite[p.~1127]{Vaidyanathan-prediction-1987} gave explicit
    polynomial predictors with bandwidth-independent endpoint error
    constants for even predictor orders. Here we obtain matching estimates
    for the optimal uniform recovery error on \([0,1]\), with comparison
    constants independent of both the sample size and the bandwidth.
    
    In Section 3 we prove the following matching bounds for the optimal uniform recovery error over the interval $[0,1]$:
	 \begin{equation}\label{onesidedoptimalbounds}
	\sqrt{\frac{\delta}{\pi}}\sin^n\frac{\delta}{2}
	\leq\rho_{n,\delta}^{-}\leq
	4\sqrt{\frac{\delta}{\pi}}\sin^n\frac{\delta}{2},
	\qquad 0<\delta<\pi,\quad n\geq1.
	\end{equation}
    Define
    \[
    c_\delta^{-}:=-\frac{\log(\sin(\delta/2))}{\pi-\delta},
    \qquad 0<\delta<\pi.
    \]
    Equivalently,
    \[
    \rho_{n,\delta}^{-}\asymp \sqrt{\frac{\delta}{\pi}}\exp\bigl(-c_\delta^-(\pi-\delta)n\bigr)
    \]
    where $c_\delta^-$ is strictly decreasing on $(0,\pi)$, $\lim_{\delta\to 0^+}c_\delta^-=+\infty$, and $
    \lim_{\delta\to\pi^-}c_\delta^-=0.$

    The bounds (\ref{onesidedoptimalbounds}) also yield a near-critical convergence criterion:
    for \(0<\delta_n<\pi\) with \(\delta_n\uparrow\pi\),
    \(\rho_{n,\delta_n}^{-}\to0\) if and only if
    \(n(\pi-\delta_n)^2\to\infty\)
    (Corollary~\ref{corollary:one sided near critical criterion}).
    At \(\delta=\pi\), \(\rho_{n,\pi}^{-}=1\) for every \(n\)
    (Proposition~\ref{proposition:one sided critical exact}), whereas
    \(\rho_{n,\pi}\asymp n^{-1/2}\).

    To compare our results with the classical work \cite{Landau-extrapolation-1986} of H. J. Landau on extrapolation of bandlimited functions, we point out that for each fixed \(0<\delta<\pi\), Landau's estimates
    \cite[Theorem~1 and Corollary]{Landau-extrapolation-1986}
    imply
    \[
    \limsup_{n\to\infty}
    \bigl(\rho_{n,\delta}^{-}\bigr)^{1/n}
    \leq \sin(\delta/2).
    \]
    Our bounds (\ref{onesidedoptimalbounds}) strengthen this fixed-bandwidth asymptotic
    upper estimate to explicit matching upper and lower bounds,
    with constants independent of both \(n\) and \(\delta\),
    for every \(n\geq1\) and \(0<\delta<\pi\).
    Landau also established exponential recovery on the growing intervals
    \[
    [-(1+\sigma)(n-1),\,\sigma(n-1)],
    \qquad
    0<\sigma<\frac{\csc(\delta/2)-1}{2}.
    \]
    As shown in the remark following
    Lemma~\ref{lemma:one sided upper bounds}, our method is able to recover this sufficient extrapolation range.

    We obtain these matching lower and upper bounds for $\rho_{n,\delta}$ and $\rho_{n,\delta}^-$ through the following  strategies:
    \begin{itemize}
        \item \emph{A common construction for the upper bounds.}
        Stieltjes representations express the transformed targets as
        integrals of resolvents. We construct polynomial approximants
        on a real interval for two-sided recovery and transfer the
        construction by a rational map to the spectral arc for one-sided
        recovery. The resulting error estimates preserve the full
        bandwidth dependence uniformly over \(x\in[0,1]\), with constants
        independent of \(n\) and \(\delta\).

        \item \emph{Matching lower bounds.}
        For two-sided recovery, a block estimate for the Legendre tail
        retains the bandwidth dependence, and bandwidth monotonicity
        extends the bound to the critical regime, giving a matching
        estimate at \(x=1/2\).
        For one-sided recovery, we construct the optimal endpoint residual
        at \(x=1\) from reversed orthogonal polynomials on the spectral arc.
        Estimating its norm gives the matching lower bound uniformly in
        \(n\) and \(\delta\).
    \end{itemize}

    We conclude the introduction with two open questions:
    \begin{enumerate}
        \item Can a numerically stable two-sided regularized Shannon sampling formula achieve an approximation error comparable to the optimal error $\rho_{n,\delta}$?

        \item Does there exist a counterpart of the regularized Shannon sampling formula for the one-sided extrapolation problem that can achieve an exponentially decaying approximation error?
    \end{enumerate}

	\section{Sharp rate for two-sided localized recovery}
	\label{section:two sided recovery}
	
	In this section we establish the sharp rate for two-sided localized recovery. Starting from the frequency-domain characterization \eqref{eq:optimal learning rate}, we first obtain an upper bound uniformly over \(x\in[0,1]\), and then derive the matching lower bound by analyzing the error at the midpoint \(x=1/2\). When \(\delta=0\), \(\mathcal B_0=\{0\}\), so both the two-sided and one-sided pointwise and uniform recovery errors vanish. We therefore restrict the proofs below to positive bandwidths; the main matching bounds remain uniform as \(\delta\downarrow0\). Throughout this section, \(n\) is a positive integer and $\delta\in(0,\pi]$.  Denote by \(\mathbb P_m\) the space of complex polynomials
	of degree at most \(m\), for integers \(m\geq0\).
	
	\subsection{Upper bound}
	We first establish an upper bound that remains uniform up to the
	critical bandwidth \(\delta=\pi\).
	For \(x\in[0,1]\), set
	\[
	\lambda:=2x-1,
	\qquad
	L:=\frac{\delta}{2}.
	\]
	Starting from \eqref{eq:optimal learning rate}, by a change of variables \(\omega=2\theta\),  we obtain
	\begin{equation}
		\begin{aligned}
			\rho_{n,\delta}(x)^2
			&=
			\frac{1}{2\pi}
			\inf_{p\in E_n}
			\int_{-\delta}^{\delta}
			\left|e^{-ix\omega}-p(\omega)\right|^2\,d\omega\\
			&=
			\frac{1}{\pi}
			\inf_{p\in E_n}
			\int_{-L}^{L}
			\left|e^{-i(\lambda+1)\theta}-p(2\theta)\right|^2\,d\theta\\
			&=
			\frac{1}{\pi}
			\inf_{p\in E_n}
			\int_{-L}^{L}
			\left|e^{-i\lambda\theta}-e^{i\theta}p(2\theta)\right|^2\,d\theta.
		\end{aligned}
		\label{eq:double-angle preliminary reduction}
	\end{equation}
	As \(p\in E_n\), \(e^{i\theta}p(2\theta)\) belongs to
	\[
	\operatorname{span}_{\mathbb C}
	\left\{
	e^{-i(2j-1)\theta}:-n+1\leq j\leq n
	\right\},
	\]
	which is the span of the odd harmonics with frequencies
	\(\pm1,\pm3,\ldots,\pm(2n-1)\).  For simplicity, set
	\[
	a_\delta:=\cos L=\cos\frac{\delta}{2},
	\qquad
	q_\delta:=\frac{1-a_\delta}{1+a_\delta}
	=\tan^2\frac{L}{2}
	=\tan^2\frac{\delta}{4}.
	\]
	For later use, also define two odd-harmonic approximation errors
	\[
	\begin{aligned}
		R_{n,L}(\lambda)
		&:=
		\inf_{c_0,\ldots,c_{n-1}\in\mathbb R}
		\biggl\|
		\cos(\lambda\theta)
		-
		\sum_{j=0}^{n-1}c_j\cos\bigl((2j+1)\theta\bigr)
		\biggr\|_{L^2([0,L])},\\
		S_{n,L}(\lambda)
		&:=
		\inf_{d_0,\ldots,d_{n-1}\in\mathbb R}
		\biggl\|
		\sin(\lambda\theta)
		-
		\sum_{j=0}^{n-1}d_j\sin\bigl((2j+1)\theta\bigr)
		\biggr\|_{L^2([0,L])}.
	\end{aligned}
	\]
	
	\begin{lemma}
		\label{lemma:double angle reduction and transfer}
		It holds for $x\in[0,1]$ that
		\begin{equation}
			\rho_{n,\delta}(x)^2
			=
			\frac{2}{\pi}
			\left(
			R_{n,L}(\lambda)^2
			+
			S_{n,L}(\lambda)^2
			\right).
			\label{eq:exact double angle reduction}
		\end{equation}
		Moreover,
		\begin{equation}
			S_{n,L}(\lambda)
			\leq
			\frac{2|\lambda|L}{\pi}\,
			R_{n,L}(\lambda).
			\label{eq:sine channel from cosine channel}
		\end{equation}
	\end{lemma}
	
	\begin{proof}
		In \eqref{eq:double-angle preliminary reduction}, the target is
		\(\cos(\lambda\theta)-i\sin(\lambda\theta)\). Its real even and
		imaginary odd parts project separately onto the cosine and sine
		channels. Symmetry of \((-L,L)\) then gives
		\eqref{eq:exact double angle reduction}.

		Let
		\[
		\mathcal C_n(\theta)=
		\sum_{j=0}^{n-1}c^{\ast}_j\cos\bigl((2j+1)\theta\bigr)
		\]
		be a best approximant of \(\cos(\lambda\theta)\) in $L^2([0,L])$, and define
		\[
		\mathcal S_n(\theta)
		:=
		\lambda\int_0^\theta \mathcal C_n(u)\,du
		=
		\sum_{j=0}^{n-1}
		\frac{\lambda c^{\ast}_j}{2j+1}
		\sin\bigl((2j+1)\theta\bigr).
		\]
		Set
		\(g(\theta):=\sin(\lambda\theta)-\mathcal S_n(\theta)\).
		Then \(g(0)=0\) and
		\(g'=\lambda(\cos(\lambda\cdot)-\mathcal C_n)\).
		The Volterra norm bound \cite[Corollary~2.2]{Ransford}, scaled to
		\((0,L)\), gives
		\[
		S_{n,L}(\lambda)\leq\|g\|_{L^2([0,L])}
		\leq\frac{2L}{\pi}\|g'\|_{L^2([0,L])}
		=\frac{2|\lambda|L}{\pi}R_{n,L}(\lambda),
		\]
		which proves \eqref{eq:sine channel from cosine channel}.
	\end{proof}
	
	The following lemma is central to our estimates. Its constant is independent of both \(n\) and \(\delta\), and the estimate remains valid at the critical bandwidth \(\delta=\pi\).
	
	\begin{lemma}[Cosine-channel estimate]
		\label{lemma:universal cosine channel estimate}
		For every \(0<\delta\leq\pi\), \(\lambda\in[-1,1]\), and
		\(n\geq1\),
		\begin{equation}
			R_{n,L}(\lambda)
			\leq
			C_{\mathrm{cos}}\,
			\frac{q_\delta^{n+1/4}}{\sqrt{n}},
			\qquad
			C_{\mathrm{cos}}
			:=
			\frac{4}{\pi}
			\left(\sqrt{\pi(2+\sqrt{2})}+\sqrt{2}\right).
			\label{eq:universal cosine channel estimate}
		\end{equation}
	\end{lemma}
	
	\begin{proof}
		Set \(\alpha:=|\lambda|\).  Since cosine is an even function, \(R_{n,L}(\lambda)=R_{n,L}(\alpha)\).  Furthermore,
		\[
		\cos\bigl((2j+1)\theta\bigr)
		=
		T_{2j+1}(\cos\theta),
		\]
		and \(T_{2j+1}\) is a Chebyshev polynomial of the first kind of
		degree \(2j+1\).  Hence there is a unique polynomial \(P_j\) of
		degree \(j\) such that
		\[
		T_{2j+1}(\cos\theta)= P_j(\cos^2\theta)\cos\theta.
		\]
		The leading coefficient of \(P_j\) is nonzero, so
		\(P_0,\ldots,P_{n-1}\) are linearly independent.  Since
		\(\dim\mathbb P_{n-1}=n\), they form a basis of
		\(\mathbb P_{n-1}\).  Consequently,
		\[
		\operatorname{span}
		\{\cos\bigl((2j+1)\theta\bigr):0\leq j\leq n-1\}
		=
		\{\cos\theta\,P(\cos^2\theta):
		P\in\mathbb P_{n-1}\}.
		\] 
		After setting \(z=\cos\theta\), we therefore
		obtain
		\begin{equation}
			R_{n,L}^2(\lambda)
			=
			\inf_{P\in\mathbb P_{n-1}}
			\int_{a_\delta}^1
			\left|\varphi_\alpha(z)-zP(z^2)\right|^2
			\frac{dz}{\sqrt{1-z^2}},
			\label{eq:cosine channel algebraic form}
		\end{equation}
		where
		\(\varphi_\alpha(z):=\cos(\alpha\arccos z)\).
		If \(\alpha=1\), then \(\varphi_\alpha(z)=z\), and the error is zero.
		We may therefore assume that \(0\le\alpha<1\).  Setting \(s=z^2\),
		we obtain
		\begin{equation}
			R^2_{n,L}(\lambda)
			=
			\inf_{P\in\mathbb P_{n-1}}
			\int_{a_{\delta}^2}^{1}
			|G_{\alpha}(s)-P(s)|^2
			\frac{\sqrt{s}}{2\sqrt{1-s}}\,ds,
			\label{eq:cosine channel algebraic form 1}
		\end{equation}
		where
		\[
		G_{\alpha}(s)
		:=
		s^{-1/2}\cos\bigl(\alpha\arccos\sqrt{s}\bigr).
		\]
		It remains to construct a suitable polynomial approximant
		\(P^{\mathrm{int}}_{n-1,\alpha}\).
		
		For \(0<\delta<\pi\), the endpoint value
		\[
		G_{\alpha}(a_{\delta}^{2})
		=
		a_{\delta}^{-1}\cos\frac{\alpha\delta}{2}
		\]
		becomes unbounded as \(\delta\uparrow\pi\). The following Stieltjes
		representation separates this endpoint behavior from the polynomial
		approximation. At \(\delta=\pi\), we use \(G_\alpha(s)\) only for
		\(s>0\), whose value at \(s=0\) does not affect the weighted
		\(L^2\) approximation.
		
		We take all square roots and inverse trigonometric functions on their
		principal branches. The function \(G_\alpha\) extends holomorphically
		to \(\mathbb C\setminus(-\infty,0]\). Across \(s>1\), the two
		boundary values of \(\arccos\sqrt{s}\) differ only by sign, so
		the even cosine removes their jump. The same evenness removes
		the local square-root branch at \(s=1\). If \(t>0\) and
		\(\eta=\operatorname{arsinh}t\), then
		\[
		\begin{aligned}
			\lim_{\varepsilon\to 0^{+}}G_\alpha(-t^2+i\varepsilon)
			&=-\frac{i}{t}
			\cos\!\left(\alpha\left(\frac\pi2-i\eta\right)\right),\\
			\lim_{\varepsilon\to 0^{+}}G_\alpha(-t^2-i\varepsilon)
			&=\frac{i}{t}
			\cos\!\left(\alpha\left(\frac\pi2+i\eta\right)\right).
		\end{aligned}
		\]
		Consequently,
		\[
		\lim_{\varepsilon\to 0^{+}}G_\alpha(-t^2+i\varepsilon)-G_\alpha(-t^2-i\varepsilon)
		=
		-\frac{2i}{t}\cos\bigl(\frac{\pi\alpha}{2}\bigr)
		\cosh(\alpha\operatorname{arsinh}t).
		\]
		Cauchy's integral formula on the keyhole contour in
		Figure~\ref{fig:keyhole-contour} gives the Stieltjes representation
		\[
		G_\alpha(s)
		=
		\int_0^\infty\frac{\sigma_\alpha(u)}{s+u}\,du,
		\qquad
		\sigma_\alpha(u)
		:=
		-\lim_{\varepsilon\to 0^{+}}\frac{G_\alpha(-u+i\varepsilon)-G_\alpha(-u-i\varepsilon)}{2\pi i}.
		\]

		\begin{figure}[htbp]
			\centering
			\includegraphics[width=0.60\linewidth]{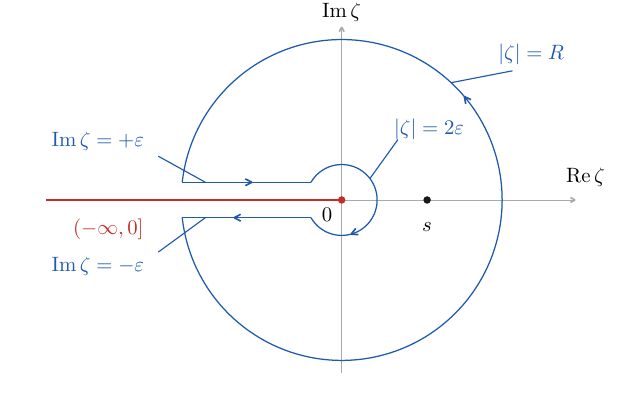}
			\caption{A positively oriented keyhole contour (blue) around
				\((-\infty,0]\) (red), with \(2\varepsilon<s<R\).}
			\label{fig:keyhole-contour}
		\end{figure}

		The formula defining the density \(\sigma_\alpha\) is the
		classical Stieltjes--Perron inversion formula
		\cite[Chapter~VIII, Theorem~7b, p.~340]{Widder-Laplace-1946}.
		Here \(\sigma_\alpha\) is locally integrable and
		\(\int_0^\infty\sigma_\alpha(u)(1+u)^{-1}\,du<\infty\).
		The small and large circles in the keyhole contour contribute zero
		because \(G_\alpha(s)=O(|s|^{-1/2})\) as \(s\to0\) and
		\(G_\alpha(s)=O(|s|^{(\alpha-1)/2})\) as \(|s|\to\infty\).
		Substituting \(u=t^2\), and putting
		\(c_\alpha:=2\cos(\pi\alpha/2)/\pi\), yields
		\begin{equation}
			G_{\alpha}(s)
			=
			c_{\alpha}\int_{0}^{\infty}
			\frac{\cosh(\alpha\operatorname{arsinh}t)}{t^2+s}\,dt.
			\label{eq:Stieltjes representation doubled cosine}
		\end{equation}
		For \(\alpha=0\), equation
		\eqref{eq:Stieltjes representation doubled cosine} reduces to
		\[
		s^{-1/2}
		=\frac{2}{\pi}\int_0^\infty\frac{dt}{s+t^2},
		\qquad s>0,
		\]
		which confirms the normalization.
		
		The affine map
		\[
		\mu(s):=\frac{2s-1-a_{\delta}^2}{1-a_{\delta}^2},
		\]
		sends \([a_{\delta}^2,1]\) onto \([-1,1]\).  For \(t\ge0\), set
		\[
		\xi(t):=\mu(-t^2)
		=-\frac{1+a_{\delta}^2+2t^2}{1-a_{\delta}^2}.
		\]
		For the Chebyshev polynomials, write
		\(\mathcal Q_{m,\delta}(s):=T_m(\mu(s))\), \(m\geq0\), and define the resolvent
		approximant
		\begin{equation}
			p^{\mathrm{int}}_{n-1,t}(s)
			:=
			\frac{1-\mathcal Q_{n,\delta}(s)/\mathcal Q_{n,\delta}(-t^2)}{s+t^2}.
			\label{eq:interval resolvent approximant}
		\end{equation}
		Because \(T_n(\mu(s))-T_n(\xi(t))\) is divisible by
		\[
		\mu(s)-\xi(t)=\frac{2(s+t^2)}{1-a_{\delta}^2},
		\]
		we have \(p^{\mathrm{int}}_{n-1,t}\in\mathbb P_{n-1}\).  Thus
		\[
		P^{\mathrm{int}}_{n-1,\alpha}(s)
		:=
		c_{\alpha}
		\int_0^\infty
		\cosh(\alpha\operatorname{arsinh}t)
		p^{\mathrm{int}}_{n-1,t}(s)\,dt
		\]
		also belongs to \(\mathbb P_{n-1}\).  Indeed, the coefficient
		integrands have finite limits at \(t=0\) by the divided-difference
		form of \eqref{eq:interval resolvent approximant}, and are integrable at
		infinity because
		\(p^{\mathrm{int}}_{n-1,t}(s)=O(t^{-2})\) coefficientwise as
		\(t\to\infty\), whereas
		\(\cosh(\alpha\operatorname{arsinh}t)=O(t^\alpha)\) and
		\(\alpha<1\).
		
		Since \(\varphi_\alpha(z)=z G_\alpha(z^2)\), subtracting
		\(zP^{\mathrm{int}}_{n-1,\alpha}(z^2)\) from the Stieltjes representation gives
		\begin{equation}
			\begin{aligned}
				\varphi_\alpha(z)-zP^{\mathrm{int}}_{n-1,\alpha}(z^2)
				&=
				z\left(G_{\alpha}(z^2)-P^{\mathrm{int}}_{n-1,\alpha}(z^2)\right)
				\\
				&=
				c_\alpha T_n(\mu(z^2))
				\int_0^{\infty}
				\frac{z\cosh(\alpha\operatorname{arsinh}t)}
				{(z^2+t^2)T_n(\xi(t))}\,dt.
			\end{aligned}
			\label{eq:explicit doubled cosine residual}
		\end{equation}
		Define
		\[
		\begin{aligned}
			\Theta_{a_{\delta}}(t)
			&:=
			\operatorname{arcosh}(-\xi(t))
			=2\operatorname{arsinh}
			\sqrt{\frac{a_{\delta}^2+t^2}{1-a_{\delta}^2}},\\
			r_{a_{\delta}}(t)
			&:=e^{-\Theta_{a_{\delta}}(t)}
			=
			\frac{\sqrt{1+t^2}-\sqrt{a_{\delta}^2+t^2}}
			{\sqrt{1+t^2}+\sqrt{a_{\delta}^2+t^2}}.
		\end{aligned}
		\]
		Then \(r_{a_{\delta}}(0)=q_{\delta}\) and
		\[
		T_n(\xi(t))=(-1)^n\cosh(n\Theta_{a_{\delta}}(t)),
		\qquad
		|T_n(\xi(t))|^{-1}\leq2r_{a_{\delta}}^n(t).
		\]
		Since \(|T_n|\leq1\) on \([-1,1]\), Minkowski's integral
		inequality, together with \eqref{eq:cosine channel algebraic form}
		and
		\eqref{eq:explicit doubled cosine residual} yields
		\begin{equation}
			\begin{aligned}
				R_{n,L}(\lambda)
				&\leq 
				\left(\int^{1}_{a_{\delta}}\frac{z^2\bigl|G_{\alpha}(z^2)-P^{\mathrm{int}}_{n-1,\alpha}(z^2)\bigr|^2}{\sqrt{1-z^2}}\,dz\right)^{1/2}
				\\
				&\leq
				c_{\alpha}\left(\int^{1}_{a_{\delta}}
				\left(\int_0^{\infty}
				\frac{1}{z^2+t^2}\frac{\cosh(\alpha\operatorname{arsinh}t)}
				{T_n(\xi(t))}\,dt\right)^2
				\frac{z^2}{\sqrt{1-z^2}}\,dz\right)^{1/2}
				\\
				&\leq
				\frac4\pi
				\int_0^\infty
				r_{a_\delta}^n(t)\sqrt{1+t^2}\,
				N_{a_\delta}(t)\,dt,
			\end{aligned}
			\label{eq:cosine channel kernel estimate}
		\end{equation}
		where we used
		\(c_\alpha\leq2/\pi\),
		\(\cosh(\alpha\operatorname{arsinh}t)\leq\sqrt{1+t^2}\), and
		\(|T_n(\xi(t))|^{-1}\leq2r_{a_{\delta}}^n(t)\).  Here
		\[
		N^2_{a_{\delta}}(t)
		:=
		\int_{a_{\delta}}^1
		\frac{z^2}{(z^2+t^2)^2\sqrt{1-z^2}}\,dz
		.
		\]
		
		Put
		\[
		M_{a_\delta}(t):=\sqrt{1+t^2}\,N_{a_\delta}(t).
		\]
		For \(t>0\), the inequality
		\(z\sqrt{1+t^2}\leq\sqrt{z^2+t^2}\), valid for \(0\leq z\leq1\),
		gives the direct estimate
		\begin{equation}
			\begin{aligned}
				N_{a_\delta}^2(t)
				&\leq
				\frac{1}{\sqrt{1+t^2}}
				\int_{a_\delta}^1
				\frac{z}{(z^2+t^2)^{3/2}\sqrt{1-z^2}}\,dz\\
				&=
				-\frac{1}{(1+t^2)^{3/2}}
				\left[\frac{\sqrt{1-z^2}}{\sqrt{z^2+t^2}}\right]_{a_\delta}^1
				=
				\frac{\sqrt{1-a_\delta^2}}
				{(1+t^2)^{3/2}\sqrt{a_\delta^2+t^2}}.
			\end{aligned}
			\label{eq:M kernel bounds}
		\end{equation}
		Since \(\sqrt{1-a_\delta^2}=(1+a_\delta)q_\delta^{1/2}\),
		we obtain
		\begin{equation}
			M_{a_\delta}(t)
			\leq
			\frac{q_\delta^{1/4}\sqrt{1+a_\delta}}
			{\bigl[(1+t^2)(a_\delta^2+t^2)\bigr]^{1/4}}.
			\label{eq:arc length kernel estimate}
		\end{equation}
		For \(a_\delta>0\), put \(u=t/a_\delta\).
		Maximizing \((1+a)/\sqrt{1+u^2a^2}\) over \(0\leq a\leq1\)
		gives
		\[
		\begin{aligned}
		\frac{(a_\delta+t)M_{a_\delta}(t)^2}{q_\delta^{1/2}}
		&\leq
		\frac{1+u}{\sqrt{1+u^2}}\,
		\frac{1+a_\delta}{\sqrt{1+u^2a_\delta^2}}\\
		&\leq
		\begin{cases}
		\dfrac{2(1+u)}{1+u^2},&0<u\leq1,\\[2mm]
		1+u^{-1},&u\geq1,
		\end{cases}
		\leq1+\sqrt{2}.
		\end{aligned}
		\]
		For \(a_\delta=0\), the same bound follows directly from
		\eqref{eq:arc length kernel estimate}. Moreover,
		\[
		\frac{t^2M_{a_\delta}(t)^2}{q_\delta^{1/2}}
		\leq
		\frac{(1+a_\delta)t^2}
		{\sqrt{(1+t^2)(a_\delta^2+t^2)}}
		\leq2.
		\]
		We therefore obtain, uniformly for \(0\leq a_\delta<1\),
		\begin{equation}
			M_{a_\delta}(t)
			\leq
			\begin{cases}
				\dfrac{\sqrt{1+\sqrt{2}}\,q_\delta^{1/4}}
				{\sqrt{a_\delta+t}},&0<t\leq1,\\[3mm]
				\sqrt{2}\,q_\delta^{1/4}t^{-1},&t\geq1.
			\end{cases}
			\label{eq:unified M kernel bound}
		\end{equation}
		
		The factor \(q_\delta^{1/4}\) in
		\eqref{eq:unified M kernel bound} retains the dependence on the
		bandwidth as \(\delta\downarrow0\). The denominator
		\(\sqrt{a_\delta+t}\) also allows \(a_\delta=0\), which is
		essential for uniformity at the critical bandwidth.

		It remains to estimate the ratio
		\[
		\frac{r_{a_{\delta}}(t)}{q_{\delta}}
		=
		\frac{r_{a_{\delta}}(t)}{r_{a_{\delta}}(0)}
		=
		e^{-\Theta_{a_{\delta}}(t)+\Theta_{a_{\delta}}(0)},
		\qquad t\geq0.
		\]
		To this end, put
		\[
		\Delta_{a_{\delta}}(t):=\Theta_{a_{\delta}}(t)-\Theta_{a_{\delta}}(0).
		\]
		Differentiating \(\Theta_{a_{\delta}}\) and integrating from \(0\)
		to \(t\) gives
		\[
		\Delta_{a_{\delta}}(t)
		=
		\int_0^t
		\frac{2v\,dv}
		{\sqrt{a_{\delta}^2+v^2}\sqrt{1+v^2}}.
		\]
		For \(0\leq t\leq1\), the inequalities
		\(\sqrt{a_{\delta}^2+v^2}\leq a_{\delta}+v\) and
		\(\sqrt{1+v^2}\leq\sqrt{2}\) imply
		\begin{equation}
			\Delta_{a_{\delta}}(t)
			\geq 
			\int_0^t
			\frac{2v\,dv}
			{\sqrt{2}(a_{\delta}+v)}
			\geq 
			\frac{\int_0^t2v\,dv}
			{\sqrt{2}(a_{\delta}+t)}
			=
			\frac{t^2}{\sqrt{2}(a_{\delta}+t)}.
			\label{eq:D small t lower}
		\end{equation}
		For \(t\geq1\), the estimate
		\[
		\frac{2v}{\sqrt{a_{\delta}^2+v^2}\sqrt{1+v^2}}
		\geq\frac1v,
		\qquad v\geq1,
		\]
		implies
		\begin{equation}
			\Delta_{a_{\delta}}(t)
			\geq 
			\Delta_{a_{\delta}}(1)+\log t
			\geq 
			\frac{1}{2\sqrt{2}}+\log t.
			\label{eq:D large t lower}
		\end{equation}

		Because \(r_{a_\delta}(t)^n=q_\delta^n
		e^{-n\Delta_{a_\delta}(t)}\), \eqref{eq:unified M kernel bound}, \eqref{eq:D small t lower} and 
		\eqref{eq:D large t lower} imply
		\[
		\begin{aligned}
			\int_0^\infty r_{a_\delta}^n(t)M_{a_\delta}(t)\,dt
			&\leq
			q_\delta^{n+1/4}
			\bigg[
			\sqrt{1+\sqrt{2}}\int_0^1
			\frac{\exp\!\left(-\frac{nt^2}{\sqrt{2}\,(a_\delta+t)}\right)}
			{\sqrt{a_\delta+t}}\,dt\\
			&\hspace{20mm}
			+\sqrt{2}\,e^{-n/(2\sqrt{2})}
			\int_1^\infty t^{-n-1}\,dt
			\bigg].
		\end{aligned}
		\]
		The substitution \(w=t/\sqrt{a_\delta+t}\) handles both
		\(a_\delta>0\) and \(a_\delta=0\) with the same Gaussian
		integral, yielding the \(n^{-1/2}\) factor with a constant 
        independent of \(\delta\).
		Indeed,
		\[
		\frac{dt}{\sqrt{a_\delta+t}}
		=
		\frac{a_\delta+t}{a_\delta+t/2}\,dw
		\leq2\,dw.
		\]
		Hence
		\[
		\int_0^1
		\frac{\exp\!\left(-\frac{nt^2}{\sqrt{2}\,(a_\delta+t)}\right)}
		{\sqrt{a_\delta+t}}\,dt
		\leq
		2\int_0^\infty e^{-nw^2/\sqrt{2}}\,dw
		=
		2^{1/4}\sqrt{\pi}\,n^{-1/2}.
		\]
		Consequently,
		\[
		\int_0^\infty r_{a_\delta}^n(t)M_{a_\delta}(t)\,dt
		\leq
		\left(\sqrt{\pi(2+\sqrt{2})}+\sqrt{2}\right)
		\frac{q_\delta^{n+1/4}}{\sqrt{n}}.
		\]
		Substitution into \eqref{eq:cosine channel kernel estimate} gives
		\[
		R_{n,L}(\lambda)
		\leq
		C_{\mathrm{cos}}\frac{q_\delta^{n+1/4}}{\sqrt{n}},
		\]
		which proves \eqref{eq:universal cosine channel estimate}.
	\end{proof}
	
	\begin{theorem}[Upper bound estimate]
		\label{theorem:upper bound estimate}
		Let
		\[
		C_{*}
		:=
		\frac{2C_{\mathrm{cos}}}{\sqrt{\pi}}
		=
		\frac{8}{\pi^{3/2}}
		\left(\sqrt{\pi(2+\sqrt{2})}+\sqrt{2}\right)
		.
		\]
		For \(0< \delta\leq\pi\) and \(n\geq1\), the uniform upper bound holds:
		\begin{equation}
			\rho_{n,\delta}
			\leq
			C_{*}\,n^{-1/2}q_\delta^{n+1/4}
			=
			\frac{C_{*}}{\sqrt{n}}\tan^{2n+1/2}\!\bigl(\frac{\delta}{4}\bigr).
			\label{eq:universal tangent rate upper bound}
		\end{equation}
	\end{theorem}
	
	\begin{proof}
		Combining Lemma~\ref{lemma:double angle reduction and transfer}
		with Lemma~\ref{lemma:universal cosine channel estimate} gives
		\[
		\begin{aligned}
			(\rho_{n,\delta}(x))^2
			&=
			\frac{2}{\pi}
			\left(R_{n,L}(\lambda)^2+S_{n,L}(\lambda)^2\right)\\
			&\leq
			\frac{2}{\pi}
			\left(1+\frac{4|\lambda|^2L^2}{\pi^2}\right)R_{n,L}(\lambda)^2\\
			&\leq
			\frac{4}{\pi}
			C_{\mathrm{cos}}^2
			\frac{q_\delta^{2n+1/2}}{n}.
		\end{aligned}
		\]
		Taking the square root on both sides proves the theorem.
	\end{proof}
	
	\subsection{Lower bound}
	
	We now derive the lower bound from the pointwise recovery error at the
	midpoint \(x=1/2\). The key is to retain a block of \(n\) terms in the
	Legendre tail. Keeping only the first term introduces an artificial
	factor that vanishes as \(\delta\uparrow\pi\).
	
	\begin{theorem}[Lower bound estimate]
		\label{theorem:lower bound estimate}
		Let
		\[
		c_*
		:=
		2^{-5/4}
		\left(\frac{1-e^{-1}}{4\pi}\right)^{1/2}.
		\]
		Uniformly for \(0<\delta\leq\pi\) and \(n\geq1\),
		\begin{equation}
			\rho_{n,\delta}
			\geq
			c_*\,n^{-1/2}q_\delta^{n+1/4}
			=
			\frac{c_*}{\sqrt{n}}\tan^{2n+1/2}(\delta/{4}).
			\label{eq:global lower bound with bandwidth factor}
		\end{equation}
	\end{theorem}
	
	\begin{proof}
		At \(x=1/2\), we have \(\lambda=0\) and
		\(S_{n,L}(0)=0\). Thus \eqref{eq:exact double angle reduction}
		and \eqref{eq:cosine channel algebraic form 1}, with
		\(A_\delta:=a_\delta^2=\cos^2(\delta/2)\), give
		\[
		\rho_{n,\delta}(1/2)^2
		=\frac{1}{\pi}\inf_{P\in\mathbb P_{n-1}}
		\int_{A_\delta}^1 |s^{-1/2}-P(s)|^2
		\sqrt{\frac{s}{1-s}}\,ds.
		\]
		We first assume that \(0<\delta<\pi\).  Since
		\[
		\sqrt{\frac{s}{1-s}}
		\geq
		\sqrt{\frac{A_{\delta}}{1-A_{\delta}}}
		=
		\cot(\delta/2),\ \ A_{\delta}\le s<1,
		\]
		\begin{equation}
			\rho_{n,\delta}(1/2)^2
			\geq
			\frac{\cot(\delta/2)}{\pi}
			\inf_{\deg P\leq n-1}
			\int_{A_\delta}^1
			\left|s^{-1/2}-P(s)\right|^2\,ds.
			\label{eq:midpoint unweighted lower reduction}
		\end{equation}
		
		Map \([A_{\delta},1]\) onto \([-1,1]\) by
		\[
		u=\frac{2s-A_{\delta}-1}{1-A_{\delta}}.
		\]
		This affine substitution maps \(\mathbb P_{n-1}\) bijectively onto
		itself.  Moreover,
		\[
		s
		=
		\cos^4(\delta/4)
		\left(1+2q_\delta u+q_\delta^2\right),
		\qquad
		ds=\frac{1-A_{\delta}}{2}\,du.
		\]
		The Legendre generating function
		\[
		(1-2zu+z^2)^{-1/2}
		=
		\sum_{k=0}^{\infty}z^kL_k(u)
		\]
		with \(z=-q_\delta\) therefore gives
		\[
		s^{-1/2}
		=
		\frac{1}{\cos^2(\delta/4)}
		\sum_{k=0}^{\infty}(-q_\delta)^kL_k(u).
		\]
		Since \(0<q_\delta<1\), the series converges uniformly on
		\([-1,1]\).  Using
		\[
		\int_{-1}^{1}L_j(u)L_k(u)\,du
		=
		\frac{2}{2k+1}\delta_{jk},
		\]
		orthogonal projection onto \(\mathbb P_{n-1}\) gives the exact
		unweighted error
		\begin{equation}
			\inf_{\deg P\leq n-1}
			\int_{A_{\delta}}^1
			\left|s^{-1/2}-P(s)\right|^2\,ds
			=
			4q_\delta
			\sum_{k=n}^{\infty}\frac{q_\delta^{2k}}{2k+1}.
			\label{eq:exact Legendre tail}
		\end{equation}
		
		Put 
		\(\tau_n(q):=\sum_{k=n}^{\infty}q^{2k}/(2k+1)\).  Since
		\[
		\cot(\delta/2)=\frac{1-q_{\delta}}{2\sqrt{q_{\delta}}},
		\]
		equations~\eqref{eq:midpoint unweighted lower reduction} and
		\eqref{eq:exact Legendre tail} imply
		\begin{equation}
			\rho_{n,\delta}(1/2)^2
			\geq
			\frac{2\sqrt{q_{\delta}}(1-q_{\delta})}{\pi}\tau_n(q_{\delta}).
			\label{eq:complete Legendre tail lower bound}
		\end{equation}
		Keeping only the first term would leave a factor
		\(1-q_\delta\) in the squared lower bound, so this estimate cannot 
        yield a positive comparison constant uniform up to \(\delta=\pi\). 
        A block of \(n\) terms cancels this loss away from a shrinking neighborhood of \(q_\delta=1\). Since \(2k+1<4n\) for \(n\leq k\leq2n-1\),
		\[
		\tau_n(q_{\delta})
		\geq
		\frac{q_{\delta}^{2n}}{4n}
		\sum_{j=0}^{n-1}q_{\delta}^{2j}
		=
		\frac{q_{\delta}^{2n}(1-q_{\delta}^{2n})}{4n(1-q_{\delta}^2)}.
		\]
		Consequently,
		\begin{equation}
			\rho_{n,\delta}(1/2)^2
			\geq
			\frac{\sqrt{q_{\delta}}}{2\pi(1+q_{\delta})}
			(1-q_{\delta}^{2n})\frac{q_{\delta}^{2n}}{n}.
			\label{eq:block Legendre tail lower bound}
		\end{equation}
		
		If \(0\le q_{\delta}\leq 1-(2n)^{-1}\), then
		\[
		q_{\delta}^{2n}
		\leq
		e^{-2n(1-q_{\delta})}
		\leq e^{-1}.
		\]
		Set \(\kappa:=\left(\frac{1-e^{-1}}{4\pi}\right)^{1/2}\).
        Since \(1+q_{\delta}\leq2\),
        \eqref{eq:block Legendre tail lower bound}
		gives
		\begin{equation}
			\rho_{n,\delta}(1/2)
			\geq
			\kappa\,n^{-1/2}q_{\delta}^{n+1/4},\quad 0\le q_{\delta}\leq 1-(2n)^{-1}.
			\label{eq:lower bound below moving threshold}
		\end{equation}
		
		In the remaining regime \(1-(2n)^{-1}<q_\delta\leq1\),
		bandwidth monotonicity extends the lower bound from the threshold 
        bandwidth uniformly to \(\delta=\pi\).
		For each fixed \(x\), the map
		\(\eta\mapsto\rho_{n,\eta}(x)\) is nondecreasing, because enlarging
		the integration interval can only increase every admissible residual
		norm.  Let \(\delta_n\in(0,\pi)\) be determined by
		\(q_{\delta_n}=1-(2n)^{-1}\).  Then \(\delta_n<\delta\), and
		\eqref{eq:lower bound below moving threshold} gives
		\[
		\rho_{n,\delta}(1/2)
		\geq
		\rho_{n,\delta_n}(1/2)
		\geq
		\kappa n^{-1/2}\left(1-\frac1{2n}\right)^{n+1/4}.
		\]
		Bernoulli's inequality yields
		\[
		\left(1-\frac1{2n}\right)^n
		\geq\frac12.
		\]
		Thus, by $1-\frac1{2n}\ge \frac12$,
		\begin{equation}
		\rho_{n,\delta}(1/2)
		\geq
		2^{-5/4}\kappa n^{-1/2}
		\geq
		c_*n^{-1/2}q_{\delta}^{n+1/4},\quad 1-(2n)^{-1}< q_{\delta}\leq 1.
        \label{eq:lower bound above moving threshold}
		\end{equation}
		Combining \eqref{eq:lower bound below moving threshold} and
		\eqref{eq:lower bound above moving threshold}, and using
		\(\rho_{n,\delta}\geq\rho_{n,\delta}(1/2)\), we obtain
		\eqref{eq:global lower bound with bandwidth factor}. 
	\end{proof}

    \subsection{Sharp estimates}
    
	Combining Theorems~\ref{theorem:upper bound estimate} and
	\ref{theorem:lower bound estimate} gives matching bounds throughout
	the bandwidth range.
	
	\begin{theorem}[Sharp rate]
		\label{theorem:sharp rate}
		For all \(0\leq\delta\leq\pi\) and \(n\geq1\), we have
		\begin{equation}
			\label{eq:sharp rate}
			\frac{c_*}{\sqrt{n}}\tan^{2n+1/2}(\delta/4)
		\leq
		\rho_{n,\delta}
		\leq
		\frac{C_*}{\sqrt{n}}\tan^{2n+1/2}(\delta/4).
		\end{equation}
	\end{theorem}

	\begin{remark}[Relation to continuous Fourier extensions]
		For \(\delta>0\), set \(t=\omega/\delta\) and
		\(T=\pi/\delta\), and define
		\[
		\mathcal G_n^{-}(T)
		:=
		\operatorname{span}
		\left\{e^{ik\pi t/T}:-n\leq k\leq n-1\right\}.
		\]
		Then the change of variables in 
		\eqref{eq:optimal learning rate} gives the identity
		\[
		\rho_{n,\delta}(x)
		=
		\sqrt{\frac{\delta}{2\pi}}
		\inf_{\phi\in\mathcal G_n^{-}(T)}
		\left\|e^{-ix\delta t}-\phi(t)\right\|_{L^2(-1,1)}.
		\]
		Thus the two-sided recovery problem is an even-dimensional continuous
		Fourier-extension problem.  If
		\[
		\mathcal G_N(T)
		:=
		\operatorname{span}\left\{e^{ik\pi t/T}:|k|\leq N\right\}
		\]
		is the standard symmetric Fourier-extension space, then
		\(\mathcal G_{n-1}(T)\subset\mathcal G_n^{-}(T)
		\subset\mathcal G_n(T)\).
		The classical extension constant
		\(E(T):=\cot^2(\pi/(4T))\) satisfies
		\[
		E(T)^{-1}
		=
		\tan^2\left(\frac{\pi}{4T}\right)
		=
		\tan^2\left(\frac{\delta}{4}\right)
		=q_\delta.
		\]
		For comparison, fix \(T>1\) and let \(\Pi_n^Tg\) be the
		\(L^2([-1,1])\) projection of an entire function \(g\) onto
		\(\mathcal G_n(T)\). The analytic-function estimate in
		\cite[Theorem~2.11]{Adcock-Huybrechs-Martin-Vaquero} gives
		\begin{equation}\label{quoted analytic-function upper estimates}
		    \|g-\Pi_n^Tg\|_{L^\infty([-1,1])}
		\leq C_{T,g}\,E(T)^{-n}.
		\end{equation}
		Related analytic-function estimates were also obtained in
		\cite[Theorem~3.1]{Webb-Coppe-Huybrechs}.
		Our two-sided bounds in Theorem~\ref{theorem:sharp rate} give an
		upper bound uniform in \(x\in[0,1]\) and a matching lower bound
		at \(x=1/2\) for the targets \(g(t)=e^{-ix\delta t}\).
		Our results give matching bounds for the worst-case scaled best \(L^2([-1,1])\)
		approximation error over \(\mathcal G^{-}_n(T)\), with the exact geometric
		factor \(E(T)^{-n}\), the algebraic prefactor \(n^{-1/2}\),
		and the bandwidth factor \(q_\delta^{1/4}\). The same comparison
		constants apply for every \(n\geq1\) and \(T\geq1\), including
		\(T=1\). The upper estimate (\ref{quoted analytic-function upper estimates}) does not
		provide the delicate matching recovery bounds (\ref{eq:sharp rate}).
	\end{remark}
	
	\section{Optimal recovery from one-sided samples}
	\label{section:one sided recovery}
	
	We now consider the extrapolation problem of reconstructing a bandlimited function $f\in \cB_\delta$ on $[0,1]$ from its sampling data at the one-sided set
	\(\{-n+1,\ldots,0\}\). For \(n\geq1\) and \(0<\delta\leq\pi\), let
	\(\Pi_{n,\delta}^-\) be the orthogonal projection in
	\(\mathcal B_\delta\), with its inherited \(L^2(\mathbb R)\) inner
	product, onto
	\[
	\mathcal V_{n,\delta}^-
	:=\operatorname{span}\{K_\delta(\cdot,j):-n+1\leq j\leq0\}.
	\]
	The projection argument in the proof of
	\cite[Theorem~2.2]{Micchelli-Xu-Zhang} applies to this sample set as well. The optimal reconstruction method is the projection $\Pi_{n,\delta}^-f$ which interpolates $f$ at \(-n+1\leq j\leq0\) and the optimal error is 
    \[
	\rho_{n,\delta}^{-}
	:=\sup_{0\leq x\le 1}\rho_{n,\delta}^{-}(x),
	\]
    where the pointwise error $\rho_{n,\delta}^{-}(x)$ is given by
    \begin{equation}
		\rho_{n,\delta}^{-}(x)
		:=
		\frac{1}{\sqrt{2\pi}}
		\inf_{c_0,\ldots,c_{n-1}\in\mathbb C}
		\biggl\|
		e^{-ix\omega}-\sum_{k=0}^{n-1}c_ke^{ik\omega}
		\biggr\|_{L^2([-\delta,\delta])},
		\qquad 0\leq x\leq1.
		\label{eq:one sided recovery error}
	\end{equation}
	Unlike two-sided recovery, one-sided recovery undergoes a phase transition at the critical bandwidth: its worst-case error tends to zero for every fixed \(0<\delta<\pi\), whereas at \(\delta=\pi\) the uniform error equals \(1\) for every \(n\).

	\subsection{The subcritical case
		\texorpdfstring{\(0<\delta<\pi\)}{0 < delta < pi}}
	
	For \(0<\delta<\pi\), set
	\[
	s_\delta:=\sin\frac{\delta}{2},
	\qquad
	a_\delta:=\cos\frac{\delta}{2},
	\qquad
	\Gamma_\delta:=
	\left\{e^{i\omega}:-\delta \le \omega\leq\delta\right\}.
	\]
	With \(z=e^{i\omega}\), the principal power satisfies
	\(z^{-x}=e^{-ix\omega}\) on \(\Gamma_\delta\).  Since \(|z|=1\) and
	\(|dz|=d\omega\),
	\begin{equation}
		\rho_{n,\delta}^{-}(x)
		=
		\frac{1}{\sqrt{2\pi}}
		\inf_{p\in\mathbb P_{n-1}}
		\left\|z^{-x}-p(z)\right\|_{L^2(\Gamma_\delta,|dz|)}.
		\label{eq:one sided arc polynomial reduction}
	\end{equation}
	Thus the one-sided problem is an algebraic polynomial-approximation
	problem on a proper circular arc.
	
	\subsubsection{Upper bound: the common resolvent construction}

	The starting point is the construction in
	\eqref{eq:Stieltjes representation doubled cosine}--\eqref{eq:interval resolvent approximant}.
	Its two steps are to represent the target as an integral of resolvents
	and to cancel each pole by a polynomial divided difference. Here, for
	\(0<x<1\), the principal branch satisfies
	\begin{equation}
		z^{-x}
		=\frac{\sin(\pi x)}{\pi}
		\int_0^\infty\frac{t^{-x}}{z+t}\,dt,
		\qquad z\in\mathbb C\setminus(-\infty,0].
		\label{eq:Stieltjes fractional power}
	\end{equation}
	The density differs from that in
	\eqref{eq:Stieltjes representation doubled cosine} while the normalized
	residual polynomial is constructed by the same procedure.

	To connect the two polynomial constructions explicitly, set
	\[
	\mathcal J(z):=\frac{(z+1)^2}{4z},\qquad z\neq0.
	\]
	Then
	\[
	\mathcal J(e^{i\omega})=\cos^2\frac{\omega}{2}\in[a_\delta^2,1],
	\qquad
	\mathcal J(-t)=-\frac{(t-1)^2}{4t}\leq0\quad(t>0).
	\]
	Thus \(\mathcal J\) maps the spectral arc to the same interval used
	in the two-sided proof, and maps the negative real axis to the
	singularity set of its Stieltjes representation. With the affine map
	\(\mu\) and the interval numerator \(\mathcal Q_{m,\delta}\) from
	\eqref{eq:interval resolvent approximant}, we have
	\[
	\zeta_\delta(z):=\mu\bigl(\mathcal J(z)\bigr)
	=\frac{z^2-2a_\delta^2z+1}{2s_\delta^2z}.
	\]
	Multiplication by \(z^m\) removes every negative power from
	\(\mathcal Q_{m,\delta}\bigl(\mathcal J(z)\bigr)\). The following construction
	realizes the resulting numerators at even degrees and supplies the
	intermediate odd degrees through a single recurrence.

	\begin{lemma}[A polynomial family on the spectral arc]
		\label{lemma:arc residual polynomials}
		Define
		\[
		F_{0,\delta}(z):=2,
		\qquad
		F_{1,\delta}(z):=\frac{z-1}{s_\delta},
		\]
		and, for \(m\geq2\),
		\begin{equation}
			F_{m,\delta}(z)
			:=
			\frac{z-1}{s_\delta}F_{m-1,\delta}(z)+zF_{m-2,\delta}(z).
			\label{eq:arc residual recurrence}
		\end{equation}
		For every \(m\geq0\), the even-degree members satisfy
		\begin{equation}
			F_{2m,\delta}(z)=2z^m\mathcal Q_{m,\delta}\bigl(\mathcal J(z)\bigr),
			\qquad z\neq0,
			\label{eq:interval arc residual identity}
		\end{equation}
		where the right-hand side extends polynomially to \(z=0\).
		Moreover, \(F_{m,\delta}\) has degree \(m\), and, for \(m\geq1\),
		\begin{equation}
			F_{m,\delta}(0)=(-s_\delta)^{-m},
			\qquad
			\max_{z\in \Gamma_\delta}|F_{m,\delta}(z)|\leq2.
			\label{eq:arc residual on arc}
		\end{equation}
		Finally, for every \(t\geq0\),
		\begin{equation}
			|F_{m,\delta}(-t)|
			\geq
			s_\delta^{-m}\max\{1,t\}^{m}.
			\label{eq:arc residual negative axis}
		\end{equation}
	\end{lemma}
	
	\begin{proof}
		Let \(\lambda_\pm(z)\) be the roots of
		\[
		\lambda^2-\frac{z-1}{s_\delta}\lambda-z=0.
		\]
		The recurrence and its two initial values give
		\[
		F_{m,\delta}(z)=\lambda_+(z)^m+\lambda_-(z)^m.
		\]
		For \(z\neq0\), their squares satisfy
		\[
		\frac{\lambda_+(z)^2+\lambda_-(z)^2}{z}=2\zeta_\delta(z),
		\qquad
		\frac{\lambda_+(z)^2\lambda_-(z)^2}{z^2}=1.
		\]
		Hence \(H_m=F_{2m,\delta}(z)/(2z^m)\), as a sequence in \(m\), has
		initial values \(H_0=1,H_1=\zeta_\delta(z)\) and satisfies the recurrence
		\[
		H_{m+1}=2\zeta_\delta(z)H_m-H_{m-1}.
		\]
		These are exactly the
		initial values and recurrence of
		\(\mathcal Q_{m,\delta}\bigl(\mathcal J(z)\bigr)\), proving
		\eqref{eq:interval arc residual identity}. The defining recurrence
		for \(F_{m,\delta}\) proves that all members are polynomials, including at
		\(z=0\).

		For \(z=e^{i\omega}\in \Gamma_\delta\), put
		\(u=\sin(\omega/2)/s_\delta\).  Then \(|u|\leq1\), and the two roots are
		\[
		\lambda_\pm(z)
		=e^{i\omega/2}
		\left(iu\pm\sqrt{1-u^2}\right).
		\]
		Both have modulus one, proving the modulus estimate in
		\eqref{eq:arc residual on arc}.  The degree assertion and the value
		at zero follow immediately from
		\eqref{eq:arc residual recurrence}.
		
		At \(z=-t\), the roots are \(-\Lambda_\delta(t)\) and \(-\frac{t}{\Lambda_{\delta}(t)}\), where
		\[
		\begin{aligned}
			\Lambda_\delta(t)
			&:=
			\frac{t+1+\sqrt{(t+1)^2-4s_\delta^2t}}{2s_\delta}.
		\end{aligned}
		\]
		Here \(\Lambda_\delta(t)>0\), and hence
		\[
		|F_{m,\delta}(-t)|=\Lambda_\delta(t)^m+\frac{t^m}{\Lambda_{\delta}(t)^m}\geq \Lambda_\delta(t)^m.
		\]
		Since
		\[
		\sqrt{(t+1)^2-4s_\delta^2t}\geq|t-1|,
		\]
		we have \(\Lambda_\delta(t)\geq s_\delta^{-1}\) for \(0\leq t\leq1\) and
		\(\Lambda_\delta(t)\geq t/s_\delta\) for \(t\geq1\).  This proves
		\eqref{eq:arc residual negative axis}.
	\end{proof}

    For \(m\geq1\) and \(t\geq0\), define
	\begin{equation}
		p^{\mathrm{arc}}_{m-1,t}(z):=\frac{1-F_{m,\delta}(z)/F_{m,\delta}(-t)}{z+t}.
		\label{eq:arc resolvent approximant}
	\end{equation}
	The numerator vanishes at \(-t\), so
	\(p^{\mathrm{arc}}_{m-1,t}\in\mathbb P_{m-1}\), and
	\[
	\frac{1}{z+t}-p^{\mathrm{arc}}_{m-1,t}(z)
	=\frac{F_{m,\delta}(z)}{(z+t)F_{m,\delta}(-t)}.
	\]
	This is the same normalized divided difference as in
    \eqref{eq:interval resolvent approximant}.
    The normalized ratio \(F_{m,\delta}(z)/F_{m,\delta}(-t)\)
    is the residual polynomial associated with this resolvent approximant.
    The arc estimate controls the numerator, and the negative-axis
    estimate controls the denominator.
	Integrating this identity against the density in
	\eqref{eq:Stieltjes fractional power} gives the upper bound below.
	
	\begin{lemma}[Upper bounds for one-sided recovery]
		\label{lemma:one sided upper bounds}
		For every \(0<\delta<\pi\), \(n\geq1\), and \(x\in[0,1]\),
		\begin{equation}
        \rho_{n,\delta}^{-}(x)
				\leq
				4\sqrt{\frac{\delta}{\pi}}\sin^n\frac{\delta}{2}
			.
			\label{eq:one sided uniform exponential estimate}
		\end{equation}
        For interior points \(0<x<1\), we also have the estimate
        \begin{equation}
            \rho_{n,\delta}^{-}(x)
				\leq
				2\sqrt{\frac{\delta}{\pi}}\sin^n\left(\frac{\delta}{2}\right)\,\Bigl(n\cos^2\frac{\delta}{2}\Bigr)^{x-1}.
                \label{eq:one sided fixed x refined estimate}
        \end{equation}
		Consequently, for each fixed \(0<\delta<\pi\), the one-sided samples
		recover \(f(x)\), uniformly for \(x\in[0,1]\), at an exponential rate.
	\end{lemma}
	
	\begin{proof}
		First let \(0<x<1\), and use the
		polynomial \(p^{\mathrm{arc}}_{n-1,t}\) of degree $n-1$ from
		\eqref{eq:arc resolvent approximant}. Define
		\begin{equation}
			P^{\mathrm{arc}}_{n-1,x}(z)
			:=
			\frac{\sin(\pi x)}{\pi}
			\int_0^\infty  p^{\mathrm{arc}}_{n-1,t}(z)t^{-x}\,dt
			\in \mathbb P_{n-1}.
			\label{eq:one sided explicit approximant}
		\end{equation}
		To justify the coefficientwise integral, write
		\[
		p^{\mathrm{arc}}_{n-1,t}(z)
		=
		\frac{F_{n,\delta}(-t)-F_{n,\delta}(z)}{(z+t)F_{n,\delta}(-t)}.
		\]
		Lemma~\ref{lemma:arc residual polynomials} implies
		\(F_{n,\delta}(-t)\neq0\) for \(t\geq0\). For fixed \(n\) and
		\(\delta\), the coefficients of \(p^{\mathrm{arc}}_{n-1,t}\) are bounded as
		\(t\downarrow0\) and are of order \(O(t^{-1})\) as
		\(t\to\infty\). The products of these coefficients with \(t^{-x}\) are therefore
		integrable at both ends when \(0<x<1\).

		Subtracting \eqref{eq:one sided explicit approximant} from
		\eqref{eq:Stieltjes fractional power} gives the exact residual
		\begin{equation}
			z^{-x}-P^{\mathrm{arc}}_{n-1,x}(z)
			=
			\frac{\sin(\pi x)}{\pi}
			F_{n,\delta}(z)
			\int_0^\infty
			\frac{t^{-x}}{(z+t)F_{n,\delta}(-t)}\,dt.
			\label{eq:one sided Stieltjes residual}
		\end{equation}
		As \(\delta\uparrow\pi\), the denominator \(|z+t|\) can be
		small near \(z=-1\), \(t=1\). To keep track of this region, we
		estimate it jointly with \(|F_{n,\delta}(-t)|\). Define
		\begin{samepage}
		\[
		D_\delta(t):=\sqrt{(t+1)^2-4s_\delta^2t}.
		\]
		Then
		\[
		\Lambda_\delta(t)=\frac{1+t+D_\delta(t)}{2s_\delta}.
		\]
		\end{samepage}
		For \(z=e^{i\omega}\in \Gamma_\delta\),
		\begin{equation}
			|z+t|^2
			=
            (t+\cos\omega)^2+\sin^2\omega
            =
            (t+1)^2-4t\sin^2\!\bigl(\frac{\omega}{2}\bigr)
			\geq D^2_\delta(t).
			\label{eq:one sided coupled denominator}
		\end{equation}
		Moreover, the negative-axis calculation in the proof of
		Lemma~\ref{lemma:arc residual polynomials} gives
		\begin{equation}
		|F_{n,\delta}(-t)|
		=\Lambda_\delta(t)^n
		+\left(\frac{t}{\Lambda_\delta(t)}\right)^n
		\geq \Lambda_\delta(t)^n.
        \label{eq:Fn estimate}
		\end{equation}
		Since \(|F_{n,\delta}(z)|\leq2\) on \(\Gamma_\delta\), it follows from
		\eqref{eq:one sided Stieltjes residual}--\eqref{eq:Fn estimate} that
		\begin{equation}
			|z^{-x}-P^{\mathrm{arc}}_{n-1,x}(z)|
			\leq
			2\frac{\sin(\pi x)}{\pi}
			\int_0^\infty
			\frac{t^{-x}}{D_\delta(t) \Lambda_\delta(t)^n}\,dt.
			\label{eq:one sided coupled residual estimate}
		\end{equation}
		
		The last integral has an exact beta-type reduction.  Set
		\[
		u=s_\delta\Lambda_\delta(t)=\frac{1+t+D_\delta(t)}{2},
		\qquad
		v=1-u^{-1}.
		\]
		As \(t\) runs from \(0\) to \(\infty\), \(u\) runs from \(1\) to
		\(\infty\), and
		\[
		t=\frac{u(u-1)}{u-s_\delta^2},
		\qquad
		\frac{dt}{D_\delta(t)}=\frac{du}{u-s_\delta^2}.
		\]
		Consequently,
		\begin{equation}
			\begin{aligned}
				\frac{\sin(\pi x)}{\pi}
				\int_0^\infty
				\frac{t^{-x}}{D_\delta(t) \Lambda_\delta(t)^n}\,dt
                &=
                s_{\delta}^n\frac{\sin(\pi x)}{\pi}
                \int_{1}^{\infty}\frac{u^{-n-x}(u-1)^{-x}}{(u-s_{\delta}^{2})^{1-x}}\,du
                \\
				&=
				s_\delta^n\frac{\sin(\pi x)}{\pi}
				\int_0^1
				\frac{v^{-x}(1-v)^{n+x-1}}
				{(a_\delta^2+s_\delta^2v)^{1-x}}\,dv\\
				&\leq
				s_\delta^na_\delta^{2x-2}
				\frac{\Gamma(n+x)}
				{\Gamma(n+1)\Gamma(x)}.
			\end{aligned}
			\label{eq:one sided moving hard edge reduction}
		\end{equation}
		The gamma ratio is bounded by
		\begin{equation}
    		\frac{\Gamma(n+x)}
    		{\Gamma(n+1)\Gamma(x)}
    		=
    		\prod_{k=1}^{n}\left(1-\frac{1-x}{k}\right)
    		\leq
    		e^{-(1-x)\sum_{k=1}^{n}k^{-1}}
    		\leq n^{x-1}.
            \label{eq:Gamma ratio estimate}
		\end{equation}
		Together with \(|\Gamma_\delta|=2\delta\), estimates \eqref{eq:one sided coupled residual estimate}--\eqref{eq:Gamma ratio estimate} yield \eqref{eq:one sided fixed x refined estimate} by
        \[
            \begin{aligned}
                \rho_{n,\delta}^{-}(x)
                &\leq\frac1{\sqrt{2\pi}}
                \|z^{-x}-P^{\mathrm{arc}}_{n-1,x}\|_{L^2(\Gamma_\delta)}\\
                &\leq\sqrt{\frac{\delta}{\pi}}\,
                \|z^{-x}-P^{\mathrm{arc}}_{n-1,x}\|_{L^\infty(\Gamma_\delta)}\\
                &\leq
                2\sqrt{\frac{\delta}{\pi}}\,
                s_\delta^n(na_\delta^2)^{x-1}.
            \end{aligned}
        \]

		We now make the bound uniform with respect to \(x\). If
		\(na_\delta^2\geq1/2\), then
		\((na_\delta^2)^{x-1}\leq2\). Hence
		\(\rho_{n,\delta}^{-}(x)\le 4\sqrt{\delta/\pi}s_\delta^n\) in this case. If
		\(na_\delta^2<1/2\), then
		\(a_\delta^2<1/2\) and
		\[
		-\log s_\delta^n
		=-\frac n2\log(1-a_\delta^2)
		\leq
		\frac{na_\delta^2}{2(1-a_\delta^2)}
		<1.
		\]
		Thus \(s_\delta^n>e^{-1}\). The zero approximant then gives
		\(\rho_{n,\delta}^{-}(x)\leq\sqrt{\delta/\pi}
		<e\sqrt{\delta/\pi}s_\delta^n\).

		This proves the bound for \(0<x<1\).
		We now treat the endpoints \(x=0\) and \(x=1\).
		At \(x=1\), the polynomial
		\[
		P^{\mathrm{arc}}_{n-1,1}(z)
		:=\frac{1-F_{n,\delta}(z)/F_{n,\delta}(0)}{z}
		=\frac{1-(-s_\delta)^nF_{n,\delta}(z)}{z}
		\]
		belongs to \(\mathbb P_{n-1}\), and
		\[
		\max_{z\in \Gamma_\delta}|z^{-1}-P^{\mathrm{arc}}_{n-1,1}(z)|
        =
        \max_{z\in \Gamma_{\delta}}
        |1-zP^{\mathrm{arc}}_{n-1,1}(z)|
		\leq2s_\delta^n.
		\]
		Hence
		\begin{equation}
		\rho_{n,\delta}^{-}(1)
		\leq2\sqrt{\frac{\delta}{\pi}}\,s_\delta^n
		.
		\label{eq:one sided endpoint upper estimate}
		\end{equation}
		At \(x=0\), take \(P^{\mathrm{arc}}_{n-1,0}(z)=1\), which has zero error.  We conclude that
		\eqref{eq:one sided uniform exponential estimate} is true.
	\end{proof}
	
    \begin{remark}[Relation to classical extrapolation estimates]
    Landau \cite{Landau-extrapolation-1986} studied exponential extrapolation
    of bandlimited functions from finite blocks of past samples.
    In our normalization, for each fixed \(0<\delta<\pi\), his results
    give exponentially accurate recovery of \(f\in\mathcal B_\delta\)
    from the data \(I_n^{-}(f)\) on the enlarged interval
    \[
        I_{n,\sigma}:=[-(1+\sigma)(n-1),\,\sigma(n-1)],\ \ 0<\sigma<b_\delta:=(\csc(\delta/2)-1)/2.
        \]
    On the fixed recovery interval \([0,1]\), his estimates also imply
    \[
    \limsup_{n\to\infty}
    \bigl(\rho_{n,\delta}^{-}\bigr)^{1/n}
    \leq \sin(\delta/2).
    \]
    Our upper bound \eqref{eq:one sided uniform exponential estimate}
    strengthens this asymptotic estimate to an explicit bound with
    constants independent of both \(n\) and \(\delta\). The matching
    lower bound is established in the next subsection.
    Below, we retain the polynomial family of
    Lemma~\ref{lemma:arc residual polynomials} and replace the Stieltjes
    representation by Cauchy's formula to recover this sufficient
    extrapolation range.
        

        For arbitrary real \(x\), we apply Cauchy's formula to the
        principal power \(z^{-x}\). On a positively oriented contour
        \(\mathscr C_R\) enclosing \(\Gamma_\delta\), with the contour and
        its interior avoiding \((-\infty,0]\) and \(F_{n,\delta}\neq0\)
        on the contour,
        the pole cancellation in \eqref{eq:arc resolvent approximant} gives
        \[
        P^{\mathrm{ext}}_{n-1,x}(z)
        :=\frac{1}{2\pi i}\int_{\mathscr C_R}
        \frac{1-F_{n,\delta}(z)/F_{n,\delta}(\zeta)}{\zeta-z}\,
        \zeta^{-x}\,d\zeta
        \in\mathbb P_{n-1}.
        \]
        The numerator vanishes at \(z=\zeta\), so the quotient has
        degree at most \(n-1\). Cauchy's formula then gives
        \[
        z^{-x}-P^{\mathrm{ext}}_{n-1,x}(z)
        =\frac{F_{n,\delta}(z)}{2\pi i}
        \int_{\mathscr C_R}
        \frac{\zeta^{-x}}{(\zeta-z)F_{n,\delta}(\zeta)}\,d\zeta.
        \]

        Since \(|F_{n,\delta}(z)|\leq2\) on \(\Gamma_\delta\), we seek a
        rational map \(\Psi_\delta\) sending the unit circle onto
        \(\Gamma_\delta\) and larger circles to enclosing contours
        \(\mathscr C_R:=\Psi_\delta(\{|w|=R\})\), \(R>1\).
        The representation in the proof of
        Lemma~\ref{lemma:arc residual polynomials} gives
        \[
        F_{n,\delta}(z)
        =\lambda_+(z)^n
        +(-1)^n\left(\frac{z}{\lambda_+(z)}\right)^n,
        \]
        for either nonzero solution \(\lambda_+(z)\) of the quadratic
        equation there. It therefore suffices to arrange
        \(\lambda_+(\Psi_\delta(w))=w\) and
        \(|\Psi_\delta(w)/w|<1\) for \(|w|>1\).
        On \(|w|=R\), the first term has modulus \(R^n\) and the second
        less than \(1\).

        As \(\theta\) varies,
        \(\operatorname{Arg}(1+s_\delta e^{i\theta})\) ranges over
        \([-\delta/2,\delta/2]\).
        Dividing \(1+s_\delta e^{i\theta}\) by its complex conjugate
        therefore doubles the argument and gives a unit-modulus map
        onto \(\Gamma_\delta\). Using \(\overline w=1/w\) on \(|w|=1\)
        gives its rational extension
        \[
        \Psi_\delta(w)
        :=\frac{1+s_\delta w}{1+s_\delta/w}
        =\frac{w(1+s_\delta w)}{w+s_\delta}.
        \]
        For \(|w|>1\), the same rational formula gives the two solutions
        \(w\) and \(-\Psi_\delta(w)/w\) of
        \(\lambda^2-(z-1)\lambda/s_\delta-z=0\) at
        \(z=\Psi_\delta(w)\). Since
        \[
        |w+s_\delta|^2-|1+s_\delta w|^2
        =(1-s_\delta^2)(|w|^2-1)>0,
        \qquad |w|>1,
        \]
        the second solution has modulus less than \(1\).
        Thus \(w\) is the unique solution outside the unit disk, and
        choosing \(\lambda_+(\Psi_\delta(w))=w\) gives
        \[
        |F_{n,\delta}(\zeta)|
        \geq R^n-\left|\frac{\zeta}{\lambda_+(\zeta)}\right|^n
        \geq R^n-1,
        \qquad \zeta\in\mathscr C_R.
        \]
        This uniqueness also makes \(\Psi_\delta\) one-to-one for
        \(|w|>1\). Its boundary correspondence and
        \(\Psi_\delta(w)\sim s_\delta w\) at infinity imply that
        \(\mathscr C_R\) is a positively oriented simple closed contour
        enclosing the arc and shrinking to it as \(R\downarrow1\); see
        Figure~\ref{fig:extrapolation-contour}.
        For \(1<R<s_\delta^{-1}\), a direct calculation gives
        \[
        m_R:=\frac{R(1-s_\delta R)}{R-s_\delta}\leq|\zeta|\leq
        M_R:=\frac{R(1+s_\delta R)}{R+s_\delta},
        \qquad \zeta\in\mathscr C_R,
        \qquad 0<m_R<1<M_R.
        \]
        The map preserves the sign of the imaginary part, and the
        contour meets the real axis only at
        \(\Psi_\delta(-R)=m_R>0\) and \(\Psi_\delta(R)=M_R>0\).
        Thus the contour and its interior avoid \((-\infty,0]\).
        \begin{figure}[htbp]
            \centering
            \includegraphics[width=0.55\linewidth]{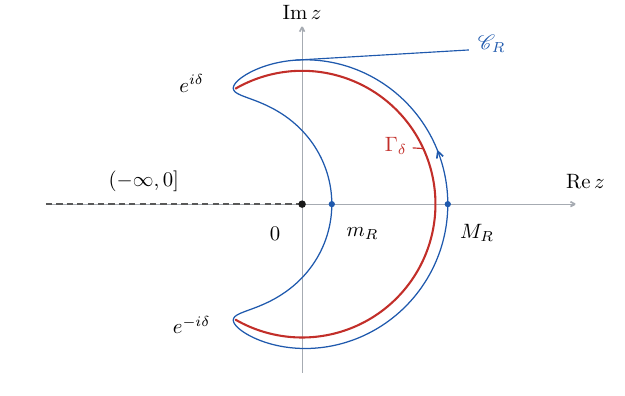}
            \caption{The spectral arc \(\Gamma_\delta\) (red) and an enclosing
                contour \(\mathscr C_R\) (blue).}
            \label{fig:extrapolation-contour}
        \end{figure}

        By the symmetry
        \(\rho_{n,\delta}^{-}(x)=\rho_{n,\delta}^{-}(-n+1-x)\),
        it suffices to consider
        \(-\sigma(n-1)\leq -x\leq(n-1)/2\).
        The error identity and \eqref{eq:one sided arc polynomial reduction},
        valid for real \(x\), then give
        \[
        \sup_{x\in I_{n,\sigma}}\rho_{n,\delta}^{-}(x)
        \leq C_{\delta,R}\sqrt{\frac{\delta}{\pi}}\,
        B_{\delta,\sigma,R}^{\,n-1},
        \qquad
        B_{\delta,\sigma,R}
        :=\frac{\max\{m_R^{-\sigma},\sqrt{M_R}\}}{R}.
        \]
        The contour has positive distance from \(\Gamma_\delta\).
        Together with \(R^n-1\geq(R-1)R^{n-1}\), this gives
        \(C_{\delta,R}>0\) independent of \(n\). Since
        \[
        \sqrt{M_R}<R,
        \qquad
        \lim_{R\downarrow1}\frac{\log R}{-\log m_R}
        =\frac{1-s_\delta}{2s_\delta}=b_\delta,
        \]
        we may fix \(R=R(\delta,\sigma)\in(1,s_\delta^{-1})\) sufficiently
        close to \(1\) so that \(B_{\delta,\sigma,R}<1\). Taking
        \(c_{\delta,\sigma}=-\log B_{\delta,\sigma,R}>0\) and
        \(C_{\delta,\sigma}=C_{\delta,R}\) proves
        \[
        \sup_{x\in I_{n,\sigma}}\rho_{n,\delta}^{-}(x)
        \leq C_{\delta,\sigma}\sqrt{\frac{\delta}{\pi}}\,
        e^{-c_{\delta,\sigma}(n-1)}.
        \]
    \end{remark}
    
	\subsubsection{Lower bound: the optimal endpoint residual}

We shall establish a matching lower bound for $\rho_{n,\delta}^-$ by examining the endpoint error $\rho_{n,\delta}^-(1)$.

	\begin{lemma}[Optimal endpoint residual and lower bound]
		\label{lemma:one sided optimal endpoint lower bound}
		Let \(0<\delta<\pi\) and \(n\geq1\). It holds
		\begin{equation}
			\rho_{n,\delta}^{-}(1)
			\geq\sqrt{\frac{\delta}{\pi}}\,s_\delta^n.
			\label{eq:one sided endpoint lower estimate}
		\end{equation}
	\end{lemma}

	\begin{proof}
    Equip \(\Gamma_\delta\)
		with the probability measure \(d\nu_\delta(z):=|dz|/(2\delta)\) and denote by \(\langle\cdot,\cdot\rangle_{\nu_\delta}\) the
		corresponding \(L^2(\nu_\delta)\) inner product. 
		Let \(\Phi_{m,\delta}\) be the degree-\(m\) monic orthogonal
		polynomial for \(\nu_\delta\), and define
		\[
		\Phi_{n,\delta}^*(z)
		:=z^n\overline{\Phi_{n,\delta}(1/\overline z)}.
		\]
		Set
		\[
		a_{m,\delta}:=\left|\Phi_{m,\delta}(0)\right|,
		\qquad m\geq1.
		\]
		 At the endpoint
		\(x=1\), the definition \eqref{eq:one sided arc polynomial reduction} of the error is equivalent to
		\[
		(\rho_{n,\delta}^{-}(1))^2
		=
		\frac{\delta}{\pi}
		\inf_{q\in 1+z\mathbb P_{n-1}}
		\|q\|_{L^2(\nu_\delta)}^2.
		\]
		Since \(\Phi_{n,\delta}\) is monic, its reversed polynomial
		\(\Phi_{n,\delta}^*\) has constant term \(1\). Moreover,
		using \(\overline z=z^{-1}\) on \(\Gamma_\delta\), orthogonality of
		\(\Phi_{n,\delta}\) gives, for \(1\leq j\leq n\),
		\[
		\left\langle\Phi_{n,\delta}^*,z^j\right\rangle_{\nu_\delta}
		=
		\overline{
		\left\langle\Phi_{n,\delta},z^{n-j}\right\rangle_{\nu_\delta}}
		=0.
		\]
		Thus \(\Phi_{n,\delta}^*\) is orthogonal to
		\(z\mathbb P_{n-1}\).  For every \(q\in1+z\mathbb P_{n-1}\), since \(\Phi_{n,\delta}^*\in 1+z\mathbb P_{n-1}\) and \(q-\Phi_{n,\delta}^*\in z\mathbb P_{n-1}\), 
		Pythagoras’ theorem therefore yields
		\[
		\|q\|_{L^2(\nu_\delta)}^2
		=
		\left\|\Phi_{n,\delta}^*\right\|_{L^2(\nu_\delta)}^2
		+
		\left\|q-\Phi_{n,\delta}^*\right\|_{L^2(\nu_\delta)}^2.
		\]
		Hence \(\Phi_{n,\delta}^*\) is the unique minimizer of	
		\[
        \min_{q\in1+z\mathbb P_{n-1}} \|q\|_{L^2(\nu_\delta)}.
        \]
		Equivalently, the optimal approximant to \(z^{-1}\) in
		\(\mathbb P_{n-1}\) is
		\[
		p_{n-1}^{\mathrm{opt}}(z)=\frac{1-\Phi_{n,\delta}^*(z)}{z}.
		\] 
        Since the
		modulus of \(\Phi_{n,\delta}^*\) agrees with that of \(\Phi_{n,\delta}\) on the unit circle,
		\(\|\Phi_{n,\delta}^*\|_{L^2(\nu_\delta)}
		=\|\Phi_{n,\delta}\|_{L^2(\nu_\delta)}\).  Finally, the
		Szeg\H{o} norm recursion
		\cite[Theorem~1.5.2 and equations~(1.5.12), (1.5.20)]{Simon-OPUC1}
		states that
		\[
		\left\|\Phi_{m,\delta}\right\|_{L^2(\nu_\delta)}^2
		=
		\left(1-\left|\Phi_{m,\delta}(0)\right|^2\right)
		\left\|\Phi_{m-1,\delta}\right\|_{L^2(\nu_\delta)}^2.
		\]
		As \(\nu_\delta\) is a probability measure,
		\(\|\Phi_{0,\delta}\|_{L^2(\nu_\delta)}=1\), and iteration gives
		\begin{equation}
			\rho_{n,\delta}^{-}(1)^2
			=\frac{\delta}{\pi}\|\Phi_{n,\delta}\|_{L^2(\nu_\delta)}^2
			=\frac{\delta}{\pi}
			\prod_{m=1}^n\bigl(1-|\Phi_{m,\delta}(0)|^2\bigr).
			\label{eq:one sided endpoint OPUC product identity}
		\end{equation}
		After rotating \(\Gamma_\delta\) by \(\pi\), it becomes the arc considered
		by Magnus with parameter \(\alpha_{\mathrm M}=1-\delta/\pi\).  The coefficient
		bound in \cite[Theorem~1.2]{Magnus-Legendre-arc} yields
		\[
		0<|\Phi_{m,\delta}(0)|<\sin\frac{\alpha_{\mathrm M}\pi}{2}
		=
		\cos\frac{\delta}{2}
		=
		a_\delta,
		\qquad m\geq1.
		\]
		Hence \(1-|\Phi_{m,\delta}(0)|^2>s_\delta^2\), and
		\eqref{eq:one sided endpoint OPUC product identity} implies
		\eqref{eq:one sided endpoint lower estimate}.
	\end{proof}

	\subsubsection{Sharp estimates}

	Combining the upper bound in Lemma~\ref{lemma:one sided upper bounds} and lower bound in Lemma~\ref{lemma:one sided optimal endpoint lower bound} gives the following theorem.

	\begin{theorem}[Sharp estimates for subcritical one-sided recovery]
		\label{theorem:one sided sharp two sided bounds}
		For every \(0<\delta<\pi\) and \(n\geq1\), the optimal uniform
		recovery error satisfies
		\begin{equation}
			\sqrt{\frac{\delta}{\pi}}\,\sin^n\frac{\delta}{2}
			\leq
			\rho_{n,\delta}^{-}
			\leq
			4\sqrt{\frac{\delta}{\pi}}\sin^n\frac{\delta}{2},
			\qquad n\geq1.
			\label{eq:one sided sharp global two sided estimate}
		\end{equation}
	\end{theorem}

	
	\begin{corollary}[Recovery as the bandwidth approaches the critical value]
		\label{corollary:one sided near critical criterion}
		Let \(0<\delta_n<\pi\) with \(\delta_n\uparrow\pi\). Then
		\[
		\rho_{n,\delta_n}^{-}\longrightarrow0
		\quad\mbox{if and only if }\quad
		n(\pi-\delta_n)^2\longrightarrow\infty.
		\]
	\end{corollary}

	\begin{proof}
		Write \(\varepsilon_n:=\pi-\delta_n\). Since
		\[
		-\log\sin\frac{\delta_n}{2}
		=-\log\cos\frac{\varepsilon_n}{2}
		=\frac{\varepsilon_n^2}{8}+O(\varepsilon_n^4),
		\]
		we have \(s_{\delta_n}^n\to0\) if and only if
		\(n\varepsilon_n^2\to\infty\). Theorem~\ref{theorem:one sided sharp two sided bounds}
		and \(\sqrt{\delta_n/\pi}\to1\) give the assertion.
	\end{proof}

	\begin{remark}[Sampling geometry and transition scales]
		To compare the two-sided and one-sided recovery problems under an equal number of sampling points, let \(N=2n\). The geometric
		factors for \(N\) two-sided and \(N\) one-sided samples are,
		respectively, \(\tan^N(\delta/4)\) and \(\sin^N(\delta/2)\).
		Their bases satisfy
		\[
		\sin\frac{\delta}{2}
		=\frac{2\tan(\delta/4)}{1+\tan^2(\delta/4)}
		>\tan\frac{\delta}{4},\qquad 0<\delta<\pi,
		\]
        which means that reconstruction from two-sided samples is potentially faster.
		Furthermore, unlike the one-sided case, the two-sided error tends to zero for
		every bandwidth sequence \(\delta_n\in[0,\pi]\), since the bound
		\(C_*n^{-1/2}\) is uniform in the bandwidth.
	\end{remark}

	\subsection{The critical case
		\texorpdfstring{\(\delta=\pi\)}{delta = pi}}
	
	At the critical bandwidth \(\delta=\pi\), $\{e^{ikt}:k\ge0\}$ is
	incomplete in \(L^2([-\pi,\pi])\). Consequently, it is not surprising to see that $\rho_{n,\pi}^-(x)$ does not tend to zero as $n\to\infty$ for any \(0<x<1\).
	
    \begin{proposition}[Exact one-sided error at the critical bandwidth]
        \label{proposition:one sided critical exact}
        For every \(n\geq1\),
        \begin{equation}
            \rho_{n,\pi}^{-}=1,
            \qquad
            \rho_{n,\pi}^{-}(0)=0,
            \qquad
            \rho_{n,\pi}^{-}(1)=1.
            \label{eq:one sided critical endpoint errors}
        \end{equation}
        For \(0<x<1\), the pointwise error satisfies
        \begin{equation}
            (\rho_{n,\pi}^{-}(x))^2
            =\frac{\sin^2(\pi x)}{\pi^2}
            \bigl[\psi_1(1-x)+\psi_1(n+x)\bigr],
            \label{eq:one sided critical trigamma error}
        \end{equation}
        where \(\psi_1(z):=\sum_{j=0}^{\infty}(z+j)^{-2}\),
        \(z>0\), is the trigamma function. Consequently,
        \begin{equation}
            \rho_{\infty,\pi}^{-}(x)
            :=\lim_{n\to\infty}\rho_{n,\pi}^{-}(x)
            =\frac{\sin \pi x}{\pi}\sqrt{\psi_1(1-x)}>0,\ \ 0<x<1.
            \label{eq:one sided critical error floor}
        \end{equation}
    \end{proposition}

    \begin{proof}
        Equip \(L^2([-\pi,\pi])\) with the normalized inner product
        \[
        \langle f,g\rangle
        :=\frac{1}{2\pi}\int_{-\pi}^{\pi}
        f(\omega)\overline{g(\omega)}\,d\omega.
        \]
        Then \(\{e^{ik\omega}:k\in\mathbb Z\}\) is an orthonormal basis,
        and the target \(e^{-ix\omega}\) has norm \(1\).
        For \(0<x<1\), the best approximant is the orthogonal 
        projection onto $\operatorname{span}\{e^{ik\omega}:0\leq k\leq n-1\}$, with
        coefficients
        \begin{equation}
                c_k^*(x)
                =\frac{1}{2\pi}\int_{-\pi}^{\pi}
                e^{-ix\omega}e^{-ik\omega}\,d\omega
                =\frac{(-1)^k\sin(\pi x)}{\pi(k+x)},
                \qquad 0\leq k\leq n-1.
            \label{eq:one sided critical coefficients}
        \end{equation}
        Pythagoras therefore gives
        \begin{equation}
            (\rho_{n,\pi}^{-}(x))^2
            =1-\frac{\sin^2(\pi x)}{\pi^2}
            \sum_{k=0}^{n-1}\frac{1}{(k+x)^2}.
            \label{eq:one sided critical finite error}
        \end{equation}
        Equivalently, Parseval's identity expresses the squared error
        as the Fourier energy in the omitted indices \(k\leq-1\) and
        \(k\geq n\). Since
        \[
        \sum_{k\leq-1}\frac{1}{(k+x)^2}=\psi_1(1-x),
        \qquad
        \sum_{k\geq n}\frac{1}{(k+x)^2}=\psi_1(n+x),
        \]
        (\ref{eq:one sided critical finite error}) gives \eqref{eq:one sided critical trigamma error}.
        Letting \(n\to\infty\) in \eqref{eq:one sided critical trigamma error} yields
        \eqref{eq:one sided critical error floor} by
        \(\psi_1(n+x)\to0\).
        At \(x=1\), the target is orthogonal to every retained mode and
        at \(x=0\), it is the first retained mode. This proves the
        endpoint values. Finally, the zero approximant gives
        \(\rho_{n,\pi}^{-}(x)\leq1\) for all \(x\in[0,1]\), while
        equality holds at \(x=1\). Hence \(\rho_{n,\pi}^{-}=1\).
    \end{proof}

	\begin{remark}[Noncommutation of limits]
		The phase transition concerns the infinite-sample limit. For every
		fixed \(n\), the recovery error is continuous in \(\delta\), whereas
		the limits \(n\to\infty\) and \(\delta\uparrow\pi\) do not commute. For
		\(0<x<1\), continuity of every finite-dimensional Gram problem in
		\(\delta\), together with the preceding results, gives
		\[
		\begin{aligned}
			\lim_{n\to\infty}\lim_{\delta\uparrow\pi}
			\rho_{n,\delta}^{-}(x)
			&=
			\frac{\sin(\pi x)}{\pi}\sqrt{\psi_1(1-x)}>0,\\
			\lim_{\delta\uparrow\pi}\lim_{n\to\infty}
			\rho_{n,\delta}^{-}(x)
			&=0.
		\end{aligned}
		\]
		Hence the limits \(n\to\infty\) and \(\delta\uparrow\pi\) do not
		commute.
	\end{remark}
	
	\section*{Data availability} 
    No datasets were generated or analysed in this theoretical study.

    \section*{Declaration of AI use} 
    The research problem and initial mathematical ideas were formulated by 
    the authors. OpenAI's ChatGPT and Codex (GPT-5.6-Sol, GPT-6-Astra and 
    GPT-6.1-Sol) were used to assist with literature searches, language 
    editing, and the exploration of certain mathematical constructions. 
    Specifically, AI tools suggested the Stieltjes-based approximation 
    construction in \eqref{eq:Stieltjes representation doubled cosine}--
    \eqref{eq:interval resolvent approximant} and assisted in adapting the 
    same strategy to the one-sided problem, leading to the polynomial family 
    in Lemma~\ref{lemma:arc residual polynomials}. The authors checked all 
    mathematical results and arguments presented in this work, reviewed and 
    revised the AI-assisted material, and take full responsibility for the 
    accuracy, originality, and integrity of the final manuscript.
    
	{\small
	\bibliographystyle{reference-numeric}
	\bibliography{ref.bib}

@article{Adcock-Huybrechs-Martin-Vaquero,
  title={On the Numerical Stability of {Fourier} Extensions},
  author={Adcock, Ben and Huybrechs, Daan and Mart{\'\i}n-Vaquero, Jes{\'u}s},
  journal={Foundations of Computational Mathematics},
  shortjournal = {Found. Comput. Math.},
  volume={14},
  number={4},
  pages={635--687},
  year={2014},
  doi={10.1007/s10208-013-9158-8}
}

@unpublished{Chen2026,
  author = {Chen, Liang and Lin, Rongrong and Zhang, Haizhang},
  title  = {Sharp Error Asymptotics and Shape-Parameter Optimization for {Shannon} Sampling with Generalized {Kaiser--Bessel} Windows},
  year   = {2026},
  month  = oct,
  doi    = {10.13140/RG.2.2.19668.16009},
  note   = {Preprint}
}

@article{Chen-Zhang-Gaussian,
  author  = {Chen, Liang and Zhang, Haizhang},
  title   = {Sharp Exponential Bounds for the {Gaussian} Regularized
             {Whittaker--Kotelnikov--Shannon} Sampling Series},
  journal = {Journal of Approximation Theory},
  shortjournal = {J. Approx. Theory},
  volume  = {245},
  pages   = {73--82},
  year    = {2019},
  doi     = {10.1016/j.jat.2019.04.003}
}

@article{Jagerman,
  author  = {Jagerman, D.},
  title   = {Bounds for Truncation Error of the Sampling Expansion},
  journal = {SIAM Journal on Applied Mathematics},
  shortjournal = {SIAM J. Appl. Math.},
  volume  = {14},
  number  = {4},
  year    = {1966},
  pages   = {714--723},
  doi     = {10.1137/0114060}
}

@article{Jiang2026,
  author  = {Jiang, Haixin and Chen, Xinyu and Chen, Liang},
  title   = {Sinh Regularized {Lagrangian} Nonuniform Sampling Series},
  journal = {Journal of Applied Mathematics and Computing},
  shortjournal = {J. Appl. Math. Comput.},
  volume  = {72},
  pages   = {67},
  year    = {2026},
  doi     = {10.1007/s12190-025-02709-4}
}

@article{Kircheis-Potts-Tasche,
  title={On Regularized {Shannon} Sampling Formulas with Localized Sampling},
  author={Kircheis, Melanie and Potts, Daniel and Tasche, Manfred},
  journal={Sampling Theory, Signal Processing, and Data Analysis},
  shortjournal = {Sampl. Theory Signal Process. Data Anal.},
  volume={20},
  number={2},
  pages={20},
  year={2022},
  publisher={Springer},
  doi={10.1007/s43670-022-00039-1}
}

@article{Kircheis2025,
  author  = {Kircheis, Melanie and Potts, Daniel and Tasche, Manfred},
  title   = {Some Remarks on Regularized {Shannon} Sampling Formulas},
  journal = {BIT Numerical Mathematics},
  shortjournal = {BIT Numer. Math.},
  volume  = {65},
  pages   = {40},
  year    = {2025},
  doi     = {10.1007/s10543-025-01081-w}
}

@article{Landau-extrapolation-1986,
  author  = {Landau, Henry J.},
  title   = {Extrapolating a Band-Limited Function from its Samples
             Taken in a Finite Interval},
  journal = {IEEE Transactions on Information Theory},
  shortjournal = {IEEE Trans. Inf. Theory},
  volume  = {32},
  number  = {4},
  pages   = {464--470},
  year    = {1986},
  doi     = {10.1109/TIT.1986.1057205}
}

@article{Magnus-Legendre-arc,
  author  = {Magnus, Alphonse P.},
  title   = {Freud Equations for {L}egendre Polynomials on a Circular Arc and
             Solution of the {G}r{\"u}nbaum--{D}elsarte--{J}anssen--{V}ries
             Problem},
  journal = {Journal of Approximation Theory},
  shortjournal = {J. Approx. Theory},
  volume  = {139},
  number  = {1--2},
  pages   = {75--90},
  year    = {2006},
  doi     = {10.1016/j.jat.2005.04.010}
}

@incollection{Micchelli-Rivlin-survey,
  author    = {Micchelli, Charles A. and Rivlin, Theodore J.},
  title     = {A Survey of Optimal Recovery},
  booktitle = {Optimal Estimation in Approximation Theory},
  editor    = {Micchelli, Charles A. and Rivlin, Theodore J.},
  series    = {The IBM Research Symposia Series},
  pages     = {1--54},
  publisher = {Springer},
  address   = {Boston, MA},
  year      = {1977},
  doi       = {10.1007/978-1-4684-2388-4_1}
}

@article{Micchelli-Xu-Zhang,
  title={Optimal learning of bandlimited functions from localized sampling},
  author={Micchelli, Charles A and Xu, Yuesheng and Zhang, Haizhang},
  journal={Journal of Complexity},
  shortjournal = {J. Complexity},
  volume={25},
  number={2},
  pages={85--114},
  year={2009},
  publisher={Elsevier},
  doi={10.1016/j.jco.2009.02.005}
}

@article{Mugler-Splettstosser-1986,
  author  = {Mugler, Dale H. and Splettst{\"o}sser, W.},
  title   = {Difference Methods for the Prediction of Band-Limited Signals},
  journal = {SIAM Journal on Applied Mathematics},
  shortjournal = {SIAM J. Appl. Math.},
  volume  = {46},
  number  = {5},
  pages   = {930--941},
  year    = {1986},
  doi     = {10.1137/0146055}
}

@article{Qian1,
  author  = {Qian, Liwen},
  title   = {On the Regularized {Whittaker--Kotel'nikov--Shannon} Sampling Formula},
  journal = {Proceedings of the American Mathematical Society},
  shortjournal = {Proc. Amer. Math. Soc.},
  volume  = {131},
  number  = {4},
  year    = {2003},
  pages   = {1169--1176},
  doi     = {10.1090/S0002-9939-02-06887-9}
}

@article{Qian2,
  author  = {Qian, Liwen and Creamer, Dennis B.},
  title   = {A Modification of the Sampling Series with a {Gaussian} Multiplier},
  journal = {Sampling Theory in Signal and Image Processing},
  shortjournal = {Sampl. Theory Signal Image Process.},
  volume  = {5},
  number  = {1},
  year    = {2006},
  pages   = {2--20},
  doi     = {10.1007/BF03549440}
}

@article{Ransford,
  author  = {Ransford, Thomas},
  title   = {Volterra Operator Norms: A Brief Survey},
  journal = {Moroccan Journal of Pure and Applied Analysis},
  shortjournal = {Moroccan J. Pure Appl. Anal.},
  volume  = {9},
  number  = {2},
  pages   = {276--290},
  year    = {2023},
  doi     = {10.2478/mjpaa-2023-0018}
}

@article{Shannon-1949,
  author  = {Shannon, Claude E.},
  title   = {Communication in the Presence of Noise},
  journal = {Proceedings of the IRE},
  shortjournal = {Proc. IRE},
  volume  = {37},
  number  = {1},
  pages   = {10--21},
  year    = {1949},
  doi     = {10.1109/JRPROC.1949.232969}
}

@book{Simon-OPUC1,
  author    = {Simon, Barry},
  title     = {Orthogonal Polynomials on the Unit Circle. {Part}~1:
               Classical Theory},
  series    = {American Mathematical Society Colloquium Publications},
  volume    = {54.1},
  publisher = {American Mathematical Society},
  address   = {Providence, RI},
  year      = {2005}
}

@article{Slepian-1965,
  author  = {Slepian, David},
  title   = {Some Asymptotic Expansions for Prolate Spheroidal Wave Functions},
  journal = {Journal of Mathematics and Physics},
  shortjournal = {J. Math. Phys.},
  volume  = {44},
  number  = {1--4},
  pages   = {99--140},
  year    = {1965},
  doi     = {10.1002/sapm196544199}
}

@article{Slepian-1976,
  author  = {Slepian, David},
  title   = {On Bandwidth},
  journal = {Proceedings of the IEEE},
  shortjournal = {Proc. IEEE},
  volume  = {64},
  number  = {3},
  pages   = {292--300},
  year    = {1976},
  doi     = {10.1109/PROC.1976.10110}
}

@article{Slepian-discrete-1978,
  author  = {Slepian, David},
  title   = {Prolate Spheroidal Wave Functions, {Fourier} Analysis,
             and Uncertainty---{V}: The Discrete Case},
  journal = {Bell System Technical Journal},
  shortjournal = {Bell Syst. Tech. J.},
  volume  = {57},
  number  = {5},
  pages   = {1371--1430},
  year    = {1978},
  doi     = {10.1002/j.1538-7305.1978.tb02104.x}
}

@article{Slepian-1983,
  author  = {Slepian, David},
  title   = {Some Comments on {Fourier} Analysis, Uncertainty and Modeling},
  journal = {SIAM Review},
  shortjournal = {SIAM Rev.},
  volume  = {25},
  number  = {3},
  pages   = {379--393},
  year    = {1983},
  doi     = {10.1137/1025078}
}

@article{Slepian-Pollak-1961,
  author  = {Slepian, David and Pollak, Henry O.},
  title   = {Prolate Spheroidal Wave Functions, {Fourier} Analysis and
             Uncertainty---{I}},
  journal = {Bell System Technical Journal},
  shortjournal = {Bell Syst. Tech. J.},
  volume  = {40},
  number  = {1},
  pages   = {43--63},
  year    = {1961},
  doi     = {10.1002/j.1538-7305.1961.tb03976.x}
}

@article{Vaidyanathan-prediction-1987,
  author  = {Vaidyanathan, P. P.},
  title   = {On Predicting a Band-Limited Signal Based on Past
             Sample Values},
  journal = {Proceedings of the IEEE},
  shortjournal = {Proc. IEEE},
  volume  = {75},
  number  = {8},
  pages   = {1125--1127},
  year    = {1987},
  doi     = {10.1109/PROC.1987.13856}
}

@article{Webb-Coppe-Huybrechs,
  author  = {Webb, Marcus and Copp{\'e}, Vincent and Huybrechs, Daan},
  title   = {Pointwise and Uniform Convergence of {Fourier} Extensions},
  journal = {Constructive Approximation},
  shortjournal = {Constr. Approx.},
  volume  = {52},
  pages   = {139--175},
  year    = {2020},
  doi     = {10.1007/s00365-019-09486-x}
}

@book{Widder-Laplace-1946,
  author    = {Widder, David Vernon},
  title     = {The {Laplace} Transform},
  series    = {Princeton Mathematical Series},
  volume    = {6},
  publisher = {Princeton University Press},
  address   = {Princeton},
  year      = {1946}
}

@article{Yang-Zhang-nonuniform,
  author  = {Yang, Yunfei and Zhang, Haizhang},
  title   = {Exponential Approximation of Band-Limited Functions from
             Nonuniform Sampling by Regularization Methods},
  journal = {Constructive Approximation},
  shortjournal = {Constr. Approx.},
  volume  = {61},
  pages   = {149--177},
  year    = {2025},
  doi     = {10.1007/s00365-024-09700-5}
}

@book{Young,
  author    = {Young, Robert M.},
  title     = {An Introduction to Nonharmonic {Fourier} Series},
  series    = {Pure and Applied Mathematics},
  volume    = {93},
  publisher = {Academic Press},
  address   = {New York},
  year      = {1980},
  isbn      = {0-12-772850-3}
}
}
\end{document}